\documentclass[11pt]{article}

\usepackage{amsfonts,amsthm}
\usepackage{amsmath}  
\usepackage{latexsym}
\usepackage{amssymb}
\usepackage{mathrsfs}
\usepackage{tikz}
\usepackage{comment}
\usepackage{graphicx}

\theoremstyle{plain}
\newtheorem{prop}{Proposition}[section]
\newtheorem{theoremm}[prop]{Theorem}

\newtheorem{lemma}[prop]{Lemma}
\newtheorem{rem}[prop]{Remark}

\theoremstyle{definition}
\newtheorem{definition}[prop]{Definition}

\numberwithin{equation}{section}

\def\be{\begin{equation}}
\def\ee{\end{equation}}
\def\bea{\begin{eqnarray}}
\def\eea{\end{eqnarray}}

\newcommand{\ie}{{\it i.e.}\ }
\newcommand{\lda}{\lambda}

\newcommand{\RR}{\mathbb{R}}
\newcommand{\ZZ}{\mathbb{Z}}
\newcommand{\CC}{\mathbb{C}}

\newcommand{\cS}{\mathcal S}
\newcommand{\cB}{\mathcal B}

\newcommand{\cK}{\mathcal K}
\newcommand{\1}{\mbox{\hspace{.0em}1\hspace{-.24em}I}}
\newcommand{\prf}{\noindent \underline{Proof:}\ }
\newcommand{\finprf}{\null \hfill {\rule{5pt}{5pt}}\\ \null}

\begin{document}

\begin{center}
	\textbf{\Large Nonlinear mirror image method for
	 nonlinear Schr\"odinger equation:\\[1ex] Absorption/emission of 
		one soliton by a boundary}\\[3ex]
	\large{Vincent Caudrelier$^a$\footnote{Corresponding author: v.caudrelier@leeds.ac.uk}, Nicolas Cramp\'e$^{b}$, Carlos Mbala Dibaya$^a$}\\[3ex]

	$^a$ School of Mathematics, University of Leeds, LS2 9JT, UK  \\[3ex]
	
	$^b$ Institut Denis-Poisson CNRS/UMR 7013 - Universit\'e de Tours - Universit\'e d’Orl\'eans, Parc de Grandmont, 37200 Tours, France.\\[3ex]

\end{center}

\vspace{0.5cm}

\begin{center} 
{ \bf Abstract}\vspace{0.5cm}\\

\begin{minipage}{15cm}

We perform the analysis of the focusing nonlinear Schr\"odinger equation on the half-line with time-dependent boundary conditions along the lines of the nonlinear method of images with the help of B\"acklund transformations. The difficulty arising from having such time-dependent boundary conditions at $x=0$ is overcome by changing the viewpoint of the method and fixing the B\"acklund transformation at infinity as well as relating its value at $x=0$ to a time-dependent reflection matrix. The interplay between the various aspects of integrable boundary conditions is reviewed in detail to brush a picture of the area. 
We find two possible classes of solutions. One is very similar to the case of Robin boundary conditions whereby solitons are reflected at the boundary, as a result of an effective interaction with their images on the other half-line. The new regime of solutions supports the existence of one soliton that is not reflected at the boundary but can be either absorbed or emitted by it. We demonstrate that this is a unique feature of time-dependent integrable boundary conditions.\\

Keywords: nonlinear Schr\"odinger equation, inverse scattering method, integrable initial-boundary value problem.
\end{minipage}
\vspace{1cm}
\end{center}

\section{Introduction}
\subsection{Integrable boundary conditions: an overview of the topic}

The problem of characterising integrable Boundary Conditions (BCs) in classical integrable systems was pioneered in \cite{sklyanin1987boundary} from the point of view of the two cornerstones of the area: its Hamiltonian formulation and its Lax pair formulation. The former views the system as a Hamiltonian field theory while the latter views it as a Partial Differential Equation (PDE). Prior to the study of integrable BCs, these two aspects had been known to go hand in hand and the natural connection between them is most readily captured by the classical $r$-matrix formalism \cite{Sklyanin:1979ye}. 
The Lax pair formulation (resp. Hamiltonian formulation) provides Lax pair integrability (resp. Liouville integrability): expanding the (trace of the) monodromy matrix $T(\lambda)$ appropriately as a series in the spectral parameter $\lambda$ one obtains a countable number of conserved quantities (resp. quantities in involution). 

These constructions guarantee integrability but say nothing about a possible solution method. The culmination of the interplay between the two formulations comes from the fact the Lax pair formulation provides a solution method known as the Inverse Scattering Method (ISM) \cite{gardner1967method} which in turns provides a solution method for the Hamiltonian formulation. From the first point of view, the ISM provides a nonlinear Fourier transform technique \cite{ablowitz1974inverse} while from the second point of view, it provides the action-angle variables in the infinite dimensional Hamiltonian setting, thus offering an infinite dimensional analog of Liouville theorem, see e.g. \cite{faddeev2007hamiltonian}.

When it comes to integrable BCs, the two aspects initially developed together in \cite{sklyanin1987boundary} have evolved rather independently since. We also note that the solution method aspect was not tackled in Sklyanin's original work. The classical Hamiltonian approach was quickly turned into its quantum counterpart \cite{sklyanin1988boundary}, as this was the original motivation for the classical $r$-matrix approach. The Lax pair point of view was developed actively in its own right along the lines of \cite{bowcock1995classically}, focusing mainly on exhibiting Lax pair integrability within integrable field theories, but not on the solution content. The Lax pair formulation as the basis for a solution method flourished independently as a means to solve initial-boundary value problems (IBVPs) following Habibullin's original work \cite{khabibullin1991backlund}. It puts forward the use of B\"acklund transformations to obtain the nonlinear analogue of the well known mirror image method for linear PDEs. The main idea is to map the half-line problem for an integrable PDE to an initial value problem on the full line solvable by the ISM. The effectiveness of the method was promptly illustrated on the nonlinear Schr\"odinger equation (NLS) in \cite{bikbaev1991initial}. This led to a wealth of subsequent activity, see for instance \cite{tarasov1991integrable,biondini2009solitons,biondini2012nonlinear,deift2011long,caudrelier2012vector,caudrelier2018interplay} for NLS, and became known as {\it nonlinear mirror image method} following the terminology of \cite{biondini2009solitons}. 
Let us note that prior to this important development, the idea of the mirror image method had been used in \cite{ablowitz1975inverse} to solve NLS with Dirichlet or Neumann BCs only, by a direct application of the odd/even extension method.

Motivated by some limitations of the nonlinear mirror image method, for instance the difficulty in dealing with dispersion relations of odd degree, as in the Korteweg-de Vries equation, Fokas and several co-workers developed another solution method, called the unified transform or Fokas method, which does not rely on mapping the problem to a full line problem, see \cite{fokas2008unified} and references therein. Instead, the idea is to perform the simultaneous spectral analysis of both parts of the auxiliary problem.

To summarize this panorama of the topic, integrable BCs have been developed along the following four directions:
\begin{enumerate}
    \item Hamiltonian formulation based on the $r$-matrix and the (classical) reflection equation, later generalised to the dynamical reflection equation \cite{sklyanin1988boundary};
    
    \item Lax pair formulation based initially on the condition $K(\lda)V(0,\lda)=V(0,-\lda)K(\lda)$, later generalised to its time-dependent version \cite{bowcock1995classically};
    
    \item Solution method 1: the nonlinear mirror image method based on Habibullin's insight;
    
    \item Solution method 2: the Fokas method, originally motivated by some limitations of the nonlinear mirror image method.
\end{enumerate}

Links between these four aspects have only been established sparsely so far. This seems to have triggered a lack of unity in the area for the last three decades and there is a tendency of confining oneself into one aspect only. The present work can be seen as a continuation of the programme initiated in \cite{ACC_2018,caudrelier2019new} whose goal is to bring unity in this topic. The situation is that we have now a clear picture of how the first three aspects naturally fit together.  

To appreciate the challenge, let us first note that, since the early days of the topic, it was believed that point 3 was different from points 1 and 2. In \cite{khabibullin1991backlund}, it is said ``we describe a method for constructing integrable initial boundary value problems which is different from the one cited in [4]'' where [4] refers to Sklyanin's work. More recently, an apparent difficulty was also noticed between points 1 and 2 and was the main motivation behind \cite{ACC_2018}. 
The following observation was made: the Hamiltonian and Lax pair approaches seem to predict different boundary matrices. Indeed, in the case of NLS, on the one hand, the relation $K(\lda)V(0,\lda)=V(0,-\lda)K(\lda)$ restricts the matrices $K(\lda)$ to be diagonal, of the form $\lda\sigma_3+i\theta\,\1$, $\theta\in\RR$. This produces the well-known Robin BCs. On the other hand, the reflection equation 
$$r_{12}(\lambda-\mu)K_1(\lambda) K_2(\mu)-   K_1(\lambda)   \ K_2(\mu) r_{21}(\lambda-\mu) +K_1(\lambda)r_{21}(\lambda+\mu) K_2(\mu)-K_2(\mu)r_{12}(\lambda+\mu) K_1(\lambda)=0\,,$$
allows for the general solution
	\begin{equation}
	K(\lda)=\lda\,A+i\theta\,\1\,,~~A\in\mathsf{sl}(2,\CC)\,,
	\end{equation}
	which leads for NLS to BCs with two arbitrary parameters, see \eqref{BCQ}, generalising the Robin case which involves only one parameter. In \cite{ACC_2018}, using results from \cite{avan2008integrable}, we resolved this apparent discrepancy at a general level and showed that the time-dependent generalisation \eqref{eq_K} of Sklyanin's condition naturally \emph{derives} from Sklyanin's Hamiltonian formulation.
	
The results provide the connection between the first and second aspects. In \cite{caudrelier2019new}, we applied the new findings to the Ablowitz-Ladik model and the explicit BCs corresponding to the more general case allowed by the reflection equation were derived. They produced new, time-dependent BCs for which the nonlinear mirror image method was implemented for the first time. This provided the connection between points 1 and 2 and the third aspect (solution method 1). In particular, this lifts the erroneous claim in \cite{khabibullin1991backlund} that the B\"acklund transformation approach is different from Sklyanin's approach. 

	The main message is as follows: the unifying equation at the junction of all approaches is
\begin{equation}
\label{key_point}
K(t,\lambda)=L(0,t,\lambda)\,.    
\end{equation}
Here $K(t,\lambda)$ is Sklyanin's reflection matrix which can depend on time in general and is a solution of \eqref{eq_K} below, and $L(x,t,\lambda)$ is the matrix realising the B\"acklund transformation in Habibullin's method. Of course, \eqref{eq_K} boils down to Sklyanin's original equation when $\partial_t K(t,\lda)=0$\footnote{We stress that $\partial_t$ means the total time derivative including the direct dependence on $t$ and the dependence on $t$ via the fields of the theory. In our example, the latter is the origin of the time dependence.}.
The relation of point 4 with the other three is only partially understood so far. In the case of a time-independent reflection matrix $K(\lda)$, the connection with points 1 and 2 is clear since Sklyanin's relation $K(\lda)V(0,\lda)=V(0,-\lda)K(\lda)$ is precisely the relation used in the Fokas method to characterise the so-called linearizable BCs. By the above discussion, since this is also the value of $L(x,t,\lda)$ at $x=0$, one can establish a connection with point 3 \ie the nonlinear mirror image method. This task was accomplished for NLS in \cite{BiondiniFD2014}. To our knowledge, this is the only study, with \cite{caudrelier2018interplay}, of the relation between the Fokas method and the nonlinear mirror image method/ISM on the full line. We stress again that this only covers the time-independent case. Therefore, from what we have explained above about the intrinsic need to allow for time-dependent BCs to unify the first and second aspects, one needs further work to connect point 4 with the unification of the first three. 

\subsection{Context of the present work}
This paper is a natural continuation of \cite{ACC_2018,caudrelier2019new} whereby we strenghten the link between the first two aspects and the third aspect of integrable BCs. Specifically, it is known that the continuous limit of the Ablowitz-Ladik model gives the NLS equation. If one computes the continuous limit of the time-dependent BCs obtained in \cite{caudrelier2019new} for the Ablowitz-Ladik model, one realises that they coincide with those derived in \cite{zambon2014classical} by a completely different method. Our main goal here is to implement the nonlinear mirror image for those time-dependent BCs for NLS. In addition to the overarching goal of our programme, this is needed for several reasons: $(i)$ this aspect was not touched in \cite{zambon2014classical}; $(ii)$ these BCs attracted attention recently \cite{gruner2020dressing,Xia:2021dca,zhang2021inverse} but the general construction of solutions (using for instance the nonlinear mirror image method) was not done. In \cite{gruner2020dressing}, only pure soliton solutions were constructed using the ``dressing the boundary'' approach \cite{Zhang_bd_dressing}. In \cite{Xia:2021dca}, the application of the nonlinear mirror image method to the new BCs was identified as an open problem that should be tackled; $(iii)$ NLS is one of the most emblematic continuous integrable models so the programme \cite{ACC_2018,caudrelier2019new} should be carried out for this system.

The paper is organised as follows. In section \ref{problem}, we formulate the initial-boundary value problem for which we will implement the nonlinear mirror image method, with emphasis on the key point \eqref{key_point}. In Section \ref{ISM}, we collect the main results of ISM for NLS that we will need. This also serves to fix notations. Section \ref{BT_review} is the main technical section which prepares the ground for a successful application of the nonlinear mirror image method with time-dependent BCs. The essential technical difference compared to the traditional implementation in \cite{bikbaev1991initial} is that the value of the B\"acklund transformation matrix at $x=0$ is now time dependent, see \eqref{key_point}, as opposed to what happens for Robin BCs. This forces us to fix its value at infinity instead. We revisit the Robin case from this point of view to make contact with the well known results in this case and help the reader identify the differences between that case and the time-dependent case. In Section \ref{BT2_section}, we show how to use the results of Section \ref{BT_review} in two steps in order to implement the nonlinear mirror image method in the time-dependent case. The main results are contained in Propositions \ref{prop_sym_discrete}, \ref{Relat_Q_QtildeBT2} and \ref{main_prop} and provide the analog for our time-dependent case of the central result \cite{bikbaev1991initial} for the Robin case. An essential difference is the possilibity for a new class of solutions that support a soliton being absorbed or emitted by the boundary. We then specialise our formulas to pure soliton solutions and recover the results of \cite{gruner2020dressing} as a particular case. We use these to visualise integrability of the time-dependent BCs by plotting a two-soliton solution being reflected at the boundary which features the graphical representation of the quantum reflection equation \cite{sklyanin1988boundary}. We also plot the new type of solutions showing a reflected soliton together with one soliton being absorbed by the boundary (see Fig.~\ref{plot_2soliton_absorbed} and \ref{plot_2soliton_absorbedb}). Conclusions and outlooks are collected in Section \ref{conclusions}. Some lengthy proofs are gathered in Appendices.

\section{Formulation of the problem: NLS with time-dependent boundary condition} \label{problem}

We consider the focusing nonlinear Schr\"odinger equation for the complex-valued function $u(x,t)$ on the half-line
\begin{eqnarray}
\label{bulk_NLS_eqs}
iu_t+u_{xx}+2 |u|^2u=0 \quad\text{for } x\ge 0,\,~~t\ge 0\,,
\end{eqnarray}
where the indices $t,x$ denote partial derivatives.
The initial data $u(x,0)=u_0(x)$ is assumed to be in the Schwartz space $\cS(\RR^+)$ and we impose the following boundary conditions (BCs) at $x=0$,
\begin{eqnarray}
\label{BCQ}
u_{xx}+(\alpha^2+ \beta^2) u \pm  2u_x\sqrt{\alpha^2-|u|^2}=0,
\end{eqnarray}
where $\alpha$ and $\beta$ are real parameters characterizing the BCs. Without loss of generality, we can fix $\alpha,\beta>0$.
Note that the BCs can equivalently be written as follows, by continuing $u$ and its derivatives to $x\to 0$ and using the bulk equation of motion,
\begin{eqnarray}
\label{timeBC}
 iu_t&=&(\alpha^2+ \beta^2) u-2|u|^2u\pm  2u_x\sqrt{\alpha^2-|u|^2} \label{BCQn}\,.
\end{eqnarray}
These BCs correspond to the third boundary condition studied in \cite{habibullin1993boundary} by a completely different method 
(the symmetry approach). We note also that, to our knowledge, the associated boundary matrix (see equation \eqref{bd_matrix_NLS} below) was derived first in \cite{zambon2014classical}, 
again using a different method (the B\"acklund transformation approach to integrable boundaries and defects). The construction of multisoliton solutions for the boundary conditions \eqref{BCQ} was initiated in \cite{gruner2020dressing,Xia:2021dca}.
From our point of view, the BCs \eqref{timeBC} represent the continuous limit of those found in \cite{caudrelier2019new}, in the same way as the NLS is the continuous limit of the Ablowitz-Ladik model.  

The bulk equation of motion and the BCs are integrable in the sense that there exits a zero curvature formulation for them. Define
\begin{eqnarray}\label{cont_pair1}
    &&U(x,t,\lambda)=\begin{pmatrix}
    -i\lambda & u(x,t)\,\\
    -u^*(x,t) & i\lambda
    \end{pmatrix}\equiv -i\lda\sigma_3+Q(x,t),\\
    \label{cont_pair2}
    &&V(x,t,\lambda)=-2i\lambda^2\sigma_3+2\lambda Q(x,t)-iQ_x(x,t)\sigma_3-iQ^2(x,t)\sigma_3,\,\\
    \label{formK}
    &&K(t,\lda)=\frac{1}{(2\lda-\beta)^2+\alpha^2}\left(-4\lda^2-4i\lda H(t)+\alpha^2+ \beta^2\right)\,,~~\label{bd_matrix_NLS}
\end{eqnarray} 
where
\begin{equation}
    \label{formH}
    H(t)=\begin{pmatrix}
    \pm\sqrt{\alpha^2-|u(0,t)|^2}&u(0,t)\\
     u^*(0,t)& \mp\sqrt{\alpha^2-|u(0,t)|^2}
    \end{pmatrix}\,.
\end{equation}
The choice of sign in \eqref{formH} corresponds to the ones in the BCs. The normalisation of $K(t,\lda)$ is chosen so that $K^{-1}(t,\lda)=K(t,-\lda)$. Note that $\sigma_3$ stands for the usual third Pauli matrix defined as $\sigma_3=\text{diag}(1,-1)$.
The asterisk denotes complex conjugate. Then, the bulk equation of motion \eqref{bulk_NLS_eqs} and the BCs \eqref{timeBC} are equivalent to
\begin{eqnarray}
    \label{eq:zcb}
    &&U_t(x,t,\lambda) -V_x(x,t,\lambda) +[U(x,t,\lambda),V(x,t,\lambda)]=0\,,\\
    \label{eq_K}
    &&K_t(t,\lambda)= V(0,t,-\lambda)  K(t,\lambda)- K(t,\lambda) V(0,t,\lambda) \,.
\end{eqnarray}
In turn, this ensures (at least locally) the existence of a $2\times 2$ matrix $\Psi(x,t,\lda)$ such that for $x\geq 0$,
\begin{eqnarray}
    \label{LP1}
    && \Psi_x(x,t,\lda)=U(x,t,\lda)\,\Psi(x,t,\lda),\\
    \label{LP2}
    && \Psi_t(x,t,\lda)=V(x,t,\lda)\,\Psi(x,t,\lda)\,,\\
    \label{LP3}
    && \Psi(0,t,-\lda)=K(t,\lda)\,\Psi(0,t,\lda)\;.
\end{eqnarray}
Eq. \eqref{LP3} signals the presence of an additional $\ZZ_2$ symmetry of the wavefunction $\Psi$ evaluated at $x=0$. Its compatibility with \eqref{LP2} continued to $x=0$ yields \eqref{eq_K}.
The connection of Eq. \eqref{LP3} with \eqref{LP1} is more subtle and is at the basis of the B\"acklund transformation approach to integrable BCs initiated in \cite{bikbaev1991initial} which we will review in section \ref{BT_review}.

\paragraph{Symmetries for the NLS equation.}
The following symmetry of the NLS Lax pair holds
\begin{equation}
    \label{reduction}
    NA(x,t,\lda^*)^*N^{-1}=A(x,t,\lda)\,,\quad\text{with}\qquad N=\begin{pmatrix}
    0 & -1\\
    1 & 0
    \end{pmatrix}\quad\text{and}\qquad A=U,V\,.
\end{equation}
The asterisk denotes (componentwise) complex conjugation. This implies in particular that the wavefunction satisfies the symmetry
\begin{equation}\label{waveFunction_NLS_symm}
    N\Psi(x,t,\lda^*)^*N^{-1}=\Psi(x,t,\lda)M(\lda)\,,
\end{equation}
where $M(\lda)$ is determined for instance by fixing the asymptotic behaviour of $\Psi$, e.g. Jost solutions. All related objects such as the scattering matrix and the B\"acklund matrices that we will use in this paper also satisfy this reduction symmetry. In particular, we have 
\begin{equation}
    NK(t,\lda^*)^* N^{-1}=K(t,\lda)\,,
\end{equation}
which boils down to $NH(t)^*N^{-1}=-H(t)$.

\paragraph{Absorption/emission of one soliton by/from the boundary.} 
Before diving into the inverse scattering method and the nonlinear mirror image method to construct solutions to the present problem on the half-line, we exhibit the simplest new solution that our results predict whereby one soliton can disappear or appear at the boundary. 
The following one-soliton solution, defined for $x,t\geq 0$,
\begin{equation}\label{eq:u1}
    u(x,t)=
    \frac{\alpha e^{i\phi} \, e^{i(\alpha^2-\beta^2)t +i\varepsilon \beta x}}
    {\cosh(\alpha(x-x_0- 2\varepsilon \beta  t)) }\,,~~\varepsilon=\pm 1\,,
\end{equation}
satisfies the focusing NLS equation and the boundary condition \eqref{BCQn}. The parameters $\phi$ and $x_0$ are the arbitrary phase and position shifts. We note that the velocity $\pm \beta$ and amplitude $\alpha$ are controlled by the boundary parameters. It was already observed in \cite{Xia:2021dca} that this solution satisfies the time-dependent boundary conditions. However the interpretation in terms of absoption or emission of a soliton was missing there. Our results show that this is the simplest instance of this phenomenon. A more complicated scenario is illustrated in Figs~\ref{plot_2soliton_absorbed} and \ref{plot_2soliton_absorbedb}. The sign $\varepsilon$ in \eqref{eq:u1} corresponds to the sign in \eqref{eq:sie} of the general construction below.
For $x_0>0$ and $\varepsilon=-1$ in \eqref{eq:u1}, the soliton disappears from the half-line $x>0$ after a time $t\sim \frac{x_0}{2\beta}$.
For $x_0<0$ and $\varepsilon=1$  in \eqref{eq:u1}, the soliton 
appears on the half-line $x>0$ after a time $t\sim -\frac{x_0}{2\beta}$. 
As we will see, this solution is a special case of 
a new class of solutions allowed by the time-dependent boundary conditions that we investigate. It breaks the intuition developed so far by the 
nonlinear mirror image method in that such a single soliton being emitted or absorbed has no mirror counterpart. 
The bulk density is defined by
\begin{equation}
\label{density}
    N(t)=\int_0^\infty |u(x,t)|^2\,dx\,.
\end{equation}
For the solution \eqref{eq:u1}, a direct calculation yields
\begin{equation}
\label{value_density}
    N(t)= \alpha  +  \alpha  \tanh( \alpha (x_0 + 2\varepsilon \beta  t ) )\,.
\end{equation}
It is clear that this is a time-dependent quantity. Its time behaviour is consistent with the picture of the one soliton being emitted or absorbed by the boundary. By this we mean, that the value of $N(t)$ as $t\to\pm\infty$ interpolates between $2\alpha$ (soliton present on the half-line) and $0$ (soliton absent on the half-line). 
Other examples of exact solutions are discussed in Section \ref{solitons}.
In particular, solutions with one soliton absorbed and one reflected are displayed in Fig.~\ref{plot_2soliton_absorbed} and \ref{plot_2soliton_absorbedb}.

\paragraph{What about integrability?}

One naturally wonders what this means given that we claim that we are dealing with (time-dependent) integrable boundary conditions, yet the bulk density \eqref{density} is not conserved in time. This is drastically different from the situation on the full line or even from the case with Robin boundary conditions. In the latter case, based on ideas in \cite{caudrelier2008systematic}, it was shown in \cite{caudrelier2012vector} that a generating function for the conserved quantities is given by, 
\begin{equation}
    I(t,\lda)=\frac{1}{2}\int_{0}^\infty u(x,t)(\Gamma(x,t,\lda)-\Gamma(x,t,-\lda))\,dx\,,
\end{equation}
where $\Gamma=\Psi_2\Psi_1^{-1}$ and $\Psi_{1,2}$ are the components of the vector solution $\Psi(x,t,\lda)=\begin{pmatrix}
    \Psi_1(x,t,\lda)\\
    \Psi_2(x,t,\lda)
\end{pmatrix}$ satisfying \eqref{LP1}-\eqref{LP3}.
Let us drop the argument $x,t,\lda$ for conciseness whenever this does not lead to confusion.

In the time-dependent case, we shall show that $I(t,\lda)$ is not conserved in time but that we can identify the correct boundary contribution which yields a generating function of conserved quantities as
\begin{eqnarray}
    \mathcal{I}(t,\lda)=I(t,\lda)-\cK(t,\lda)\,,
\end{eqnarray}
where
\begin{equation}
\label{def_bd_charges}
    \cK(t,\lda)=\frac{1}{2}\ln\left(K_{11}(t,\lda)+K_{12}(t,\lda)\Gamma(0,t,\lda)\right)\,.
\end{equation}
The proof goes as follows. 
A direct calculation using \eqref{LP1}-\eqref{LP2} yields
\begin{equation}
\label{conservation}
    \partial_t (u\Gamma)=\partial_x(V_{11}+V_{12}\Gamma)\,.
\end{equation}
This can be used for $\Gamma(x,t,\lda)$ and for $\Gamma(x,t,-\lda)$ to yield
\begin{equation}
    \partial_t I(t,\lda)=\frac{1}{2}\left(-(V_{11}(0,t,\lda)+V_{12}(0,t,\lda)\Gamma(0,t,\lda))+(V_{11}(0,t,-\lda)+V_{12}(0,t,-\lda)\Gamma(0,t,-\lda))\right)\,.
\end{equation}
We now use \eqref{LP3} to obtain 
$$\Gamma(0,t,-\lda)=(K_{21}(t,\lda)+K_{22}(t,\lda)\Gamma(0,t,\lda))(K_{11}(t,\lda)+K_{12}(t,\lda)\Gamma(0,t,\lda))^{-1}$$
and we use \eqref{eq_K} to eliminate $V_{11}(0,t,-\lda)$ and $V_{12}(0,t,-\lda)$. We get, after some cancellations
\begin{eqnarray}
    &&\partial_t I(t,\lda)=\frac{1}{2}\left(  \partial_tK_{11}(t,\lda)+\partial_tK_{12}(t,\lda)\Gamma(0,t,\lda)\right)(K_{11}(t,\lda)+K_{12}(t,\lda)\Gamma(0,t,\lda))^{-1}\\\nonumber
    &&+\frac{1}{2} K_{12}(t,\lda) \left(V_{21}(0,t,\lda)-2\lda V_{11}(0,t,\lda)-V_{12}(0,t,\lda)\Gamma^2(0,t,\lda)\right)(K_{11}(t,\lda)+K_{12}(t,\lda)\Gamma(0,t,\lda))^{-1}\,.
\end{eqnarray}
It remains to note that \eqref{cont_pair2} implies the following Riccati equation in time for $\Gamma$
$$V_{21}(0,t,\lda)-2\lda V_{11}(0,t,\lda)-V_{12}(0,t,\lda)\Gamma^2(0,t,\lda)=\partial_t \Gamma(0,t,\lda)\,.$$
With this, we deduce
\begin{eqnarray}
    \partial_t I(t,\lda)= \frac{1}{2}\partial_t\ln\left(K_{11}(t,\lda)+K_{12}(t,\lda)\Gamma(0,t,\lda)\right)\,.
\end{eqnarray}
This shows that $\partial_t I(t,\lda)\neq 0$ but leads naturally to introduce $\cK(t,\lda)$ as in \eqref{def_bd_charges} and
\begin{equation}
    \label{modif_conservation}
\partial_t {\cal I}(t,\lda)=\partial_t \left( I(t,\lda)-\cK(t,\lda)\right)=0  \,,
\end{equation}
which shows the announced result.

Let us remark that for Robin boundary conditions, since $K$ is diagonal and time independent (see Eq. \eqref{eq:Krobin}), $K_{12}(t,\lda)=0$ and $K_{11}(t,\lda)=\lda$, the previous equation simplifies and shows $I(t,\lda)$
is the generating series for an infinite numbers of conserved quantities without needing $\cK(t,\lda)$. However, for the time-dependent boundary conditions we study here, this is not the case and $\cK(t,\lda)$ is indeed time-dependent and exactly compensates for the loss of conservation in time of $I(t,\lda)$. 
Integrability holds for the system ``half-line+boundary'' while the half-line only can be thought as being an open system coupled to a boundary that acts as a reservoir. 

We should clarify how to compute the conserved quantities 
which are the coefficients of the generating series ${\cal I}(t,\lda)=\sum_{n=0}^\infty\frac{{\cal I}_n(x,t)}{(2i\lda)^n}$. This
computation is based on the fact that
$\Gamma$ satisfies the following Riccati equation 
$$\Gamma_x=-u^*+2i\lda\Gamma-u\Gamma^2\,,$$
and it admits a series expansion in $1/\lda$,
$$\Gamma(x,t,\lda)=\sum_{n=1}^\infty\frac{\Gamma_n(x,t)}{(2i\lda)^n}\,.$$
This allows one to obtain the quantities $I_{2n-1}=\int_{0}^\infty (u\Gamma_{2n-1})\,dx $ recursively from 
$$\Gamma_1=u^*\,,~~\Gamma_{n+1}=(\Gamma_{n})_x+u\sum_{k=1}^{n-1}\Gamma_k\Gamma_{n-k}\,.$$
This gives $I_1=\int_{0}^\infty |u(x,t)|^2\,dx$ which is not conserved on its own from \eqref{value_density} as mentioned previously.
However, our formula \eqref{modif_conservation} tells us that the combination\footnote{Here and in the next formula, we have dropped irrelevant constants related to the normalisation of the matrix $K$.}
\begin{equation}
\label{I1}
 {\cal I}_1=  I_1-\cK_1  =\int_{0}^\infty |u(x,t)|^2 dx \pm \sqrt{\alpha^2-|u(0,t)|^2}
\end{equation}
will be conserved in time. For our example \eqref{eq:u1}, we have $I_1=N(t)$ found in \eqref{value_density}. Noticing that the $\pm$ sign 
in front of the square root is equal to\footnote{This can be found by checking the boundary condition \eqref{BCQ} for solution \eqref{eq:u1}.} $-\text{sign}(x_0+2\varepsilon\beta t)$ we have $\pm\sqrt{\alpha^2-|u(0,t)|^2}=-\alpha\tanh(\alpha(x_0+2\varepsilon \beta t))$, hence the result.
Similarly, the Hamiltonian of this system can be recognized in ${\cal I}_3$ and can be computed exactly
\begin{equation}
\label{I3}
   {\cal I}_3=- \int_{0}^\infty\left( |u_x(x,t)|^2 - |u(x,t)|^4\right) dx
   \mp\frac{3\beta^2-\alpha^2-2|u(0,t)|^2}{3}\sqrt{\alpha^2-|u(0,t)|^2} \,.
\end{equation}
Again, for solution \eqref{eq:u1}, a direct computation gives
\begin{equation}
    \int_{0}^\infty\left( |u_x(x,t)|^2 - |u(x,t)|^4\right) dx=\alpha\tanh{(\alpha(x_0+2\varepsilon \beta t))}\frac{3\beta^2-\alpha^2-2|u(0)|^2}{3}
\end{equation}
and conservation in time of ${\cal I}_3$ holds using again the value of $\pm\sqrt{\alpha^2-|u(0,t)|^2}$ given above. Note that expressions \eqref{I1} and \eqref{I3} were obtained in \cite{zambon2014classical} using Sklyanin's double-row formalism. However, the generating function ${\cal K}$ for the boundary contribution was not identified there. 

\section{Review of the Inverse Scattering Method (ISM)}\label{ISM}

The nonlinear mirror image approach to initial-boundary value problems relies on the use of ISM, initially developed to construct solutions of Initial Value Problems (IVPs) for integrable nonlinear PDEs. Therefore, in this section we briefly outline the ingredients we need from ISM applied to the NLS equation \eqref{bulk_NLS_eqs} with the initial condition in $\mathcal{S}(\mathbb{R})$. More general functional spaces are possible but our aim in this paper is to focus on methods and ideas rather than on technicalities related to functional spaces. 
The basis of the method relies on the spectral analysis of \eqref{LP1} at $t=0$ to get the initial scattering data (direct part).
The key point is that, under the time evolution induced by \eqref{LP2}, the time evolution of the scattering data is linear. The inverse part consist in using the time evolved scattering data in the reconstruction formula for the solution of NLS.

Specifically, hereafter $\mathbb{C}^+$ and $\mathbb{C}^-$ are the upper and lower complex plane respectively, that is, $\mathbb{C}^+=\{\lda\in \mathbb{C}\,\rvert\, \text{Im}(\lda)>0\}$, $\mathbb{C}^-=\{\lda\in \mathbb{C}\,\rvert\, \text{Im}(\lda)<0\}$, and $\overline{\mathbb{C}^\pm}$ is the closure of $\mathbb{C}^\pm$ containing the real line. 

\subsection{Direct part}
Given the initial data $u(x,t=0)=u(x)\in\cS(\mathbb{R})$, the construction of the scattering data is achieved by considering the unique $2\times 2$ fundamental matrix solutions of \eqref{LP1} at $t=0$, denoted by $\Psi_\pm(x,\lda)$ and called Jost solutions, which satisfy (see e.g. \cite{ablowitz1975inverse,ablowitz1981solitons})
\begin{equation}\label{asymp_JostSols}
	\lim_{x\to\pm\infty}\Psi_\pm(x,\lda)e^{i\lda x\sigma_3}=\1, \quad \lda\in \mathbb{R}.
\end{equation}
We denote by $\Psi_\pm^{(1)}(x,\lda)$ and $\Psi_\pm^{(2)}(x,\lda)$ the first and second column of $\Psi_\pm(x,\lda)$, respectively. It can be shown (see, e.g. \cite{ablowitz1974inverse,ablowitz1981solitons,beals1984scattering,petermiller2}) that $\Psi_\pm^{(1)}(x,\lda)$ and $\Psi_\mp^{(2)}(x,\lda)$ 
are continuous on $\overline{\mathbb{C}^\mp}$ and have an analytic continuation in $\mathbb{C}^\mp$.
From \eqref{LP1} and (\ref{asymp_JostSols}), one deduces that $\Psi_+(x,\lda)$ and $\Psi_-(x,\lda)$ have determinant equal to $1$ and hence are two \emph{fundamental solutions} of \eqref{LP1} for $\lda\in \mathbb{R}$. Therefore, they must be related as 
\begin{equation}\label{rel_S}
    \Psi_+(x,\lda)=\Psi_-(x,\lda)\,S(\lda)\,,~~\lda\in\RR\,,
\end{equation}
where the matrix $S(\lda)$ is called the \emph{scattering matrix} associated to $u(x)$ or $Q(x)=\begin{pmatrix}
    0&u(x)\\-u^*(x)&0
\end{pmatrix}$. The wavefunctions $\Psi_\pm(x,\lda)$ satisfy the symmetry relation \eqref{waveFunction_NLS_symm} with $M=\1$. We deduce that
$\Psi_\pm(x,\lda)^*=N^{-1}\Psi_\pm(x,\lda)N$ and
\begin{equation}
	S(\lda)=\begin{pmatrix}
		a^*(\lda) & - b^*(\lda)\\
		b(\lda) & a(\lda)
	\end{pmatrix}, \quad \lda\in \mathbb{R},
\end{equation}
where $a(\lda)$ and $b(\lda)$ are complex-valued functions defined on $\mathbb{R}$.
The matrix $S(\lda)$ is uni-modular, that is $\det S(\lda)=1$. Explicitly, we have $|a(\lda)|^2+|b(\lda)|^2=1, \lda\in \mathbb{R}.$
From \eqref{rel_S} it follows
\begin{equation}\label{a_Jost_sols}
    a(\lda)=\det\left(\Psi_-^{(1)}(x,\lda),\Psi_+^{(2)}(x,\lda)\right),\quad a^*(\lda)=-\det\left(\Psi_-^{(2)}(x,\lda),\Psi_+^{(1)}(x,\lda)\right).
\end{equation}
Hence $a(\lda)$ (resp. $a^*(\lda)$) is continuous on $\overline{\mathbb{C}^+}$ (resp. $\overline{\mathbb{C}^-}$) and has an analytic continuation on $\mathbb{C}^+$ (resp. $\mathbb{C}^-$). Note that in this setting (that is, $u\in\mathcal{S}(\mathbb{R})$) $b(\lda)$ is continuous on $\mathbb{R}$ as well, but cannot be continued analytically into any region of $\mathbb{C}$. Hereafter, whenever we mention any column of Jost matrix solutions $\Psi_\pm(x,\lda)$ (or $a(\lda)$, $a^*(\lda)$), we always refer to their analytic continuation on the appropriate region of the complex plane. 

The continuous scattering coefficient $a(\lda)$ may have zeroes in $\mathbb{C}^+$. In such a case, we will make the standard assumption that there is a finite number $n$ of simple zeros $\lda_k$, $k=1,\dots, n$. Following the terminology used in \cite{beals1984scattering}, potentials $u(x)$ that lead to such a property of the continuous scattering coefficient $a(\lda)$ are called \emph{generic potentials}. In what follows, all potentials are assumed to be generic.  Thus the first equation in \eqref{a_Jost_sols} implies $\Psi_-^{(1)}(x,\lda_k)=\gamma(\lda_k)\Psi_+^{(2)}(x,\lda_k),$
where the proportionality constants $\gamma(\lda_k)$ are called \emph{norming constants}. The quantities $c(\lda_k)=\frac{\gamma(\lda_k)}{a^\prime(\lda_k)}$, $k=1,\dots,n$, are sometimes used instead of $\gamma(\lda_k)$ and are also referred to as norming constants. We will use both depending on our needs. The zeroes and norming constants together form the so-called \emph{discrete scattering data}.
The \emph{reflection coefficient}, denoted $r(\lda)$, is a complex-valued function defined as $r(\lda)=\frac{b(\lda)}{a(\lda)}$ for $\lda\in \mathbb{R}$. It has been shown in \cite{beals1984scattering} that $r(\lda)\in \mathcal{S}(\mathbb{R})$ with the property $\|r\|_\infty<1$. The discrete scattering data and the reflection coefficient form the \emph{scattering data} necessary for the direct and inverse part of the method as we now recall.
\begin{definition}\label{s-map}
  The map $S$ associates to $u\in \mathcal{S}(\mathbb{R})$ its scattering data, namely
  $$\begin{array}{cccc}
S: &     \mathcal{S}(\mathbb{R}) & \longrightarrow & \mathcal{S}(\mathbb{R})\times (\mathbb{C}^+)^n\times (\mathbb{C})^n\\
    &  u & \longmapsto  & \left(r,\lda_1,\dots,\lambda_n,c(\lda_1),\dots,c(\lda_n)\right)
  \end{array}$$ 
\end{definition}

\subsection{Inverse part}
The inverse of the map $S$ can be obtained via the following normalized Riemann-Hilbert problem (see e.g. \cite{deift2019riemannhilbert} for details). Given $\left(r,\lda_1,\dots,\lambda_n,c(\lda_1),\dots,c(\lda_n)\right)$,
construct the (unique) $2\times 2$ matrix function $\mu(x,\lda)$ which satisfies the following properties: 
\begin{itemize}
    \item {\bf Analyticity.} $\mu(x,\lda)$ is analytic in $\mathbb{C}\backslash\left( \mathbb{R}\cup\left\{\lda_1,\dots,\lda_n,\lda_1^*,\dots,\lda_n^*\right\}\right)$;
    
    \item {\bf Jump condition.} It has continuous boundary values on the real line $\displaystyle \mu_\pm(x,\lda):=\lim_{\varepsilon\to0^+}\mu(x,\lda\pm i\varepsilon)$, $\lda\in \mathbb{R}$, satisfying the \emph{jump condition}
    \begin{equation*}
        \mu_+(x,\lda)=\mu_-(x,\lda)v(x,\lda),\, \lda\in \mathbb{R},
    \end{equation*}
    where the \emph{jump matrix} is given by
    \begin{equation*}
	    v(x,\lda):=\begin{pmatrix}
		1+|r(\lda)|^2 & -r^*(\lda)e^{-2i\lda x}\\
		-r(\lda)e^{2i\lda x} & 1
	    \end{pmatrix};
    \end{equation*}
    \item {\bf Residues.} $\mu(x,\lda)$ has simple poles at $\lda_k,\lda_k^*$ for $k=1,\dots,n$, and the residues are given by
    \footnotesize
    \begin{equation}\label{residues_mu}
	    \underset{\lda=\lda_k}{\text{Res}}\mu(x,\lda)=\lim_{\lda\to \lda_k}\mu(x,\lda)\begin{pmatrix}
		0&0\\ c(\lda_k)e^{2i\lda_k x}&0
	    \end{pmatrix},\, \underset{\lda=\lda_k^*}{\text{Res}}\mu(x,\lda)=\lim_{\lda\to \lda_k^*}\mu(x,\lda)\begin{pmatrix}
		0&-c^*(\lda_k)e^{-2i\lda_k^*x}\\ 0&0
	    \end{pmatrix};
    \end{equation}
    \normalsize
    \item {\bf Normalisation.} $\mu(x,\lda)\to\1$ as $\lda\to\infty$.
\end{itemize}
\begin{definition}\label{inverse-s-map}
    The map $P$ takes $\left(r,\lda_1,\dots,\lambda_n,c(\lda_1),\dots,c(\lda_n)\right)\in \mathcal{S}(\mathbb{R})\times (\mathbb{C}^+)^n\times (\mathbb{C})^n$ to
    \begin{equation*}
        u(x)=2i(\mu_1(x))_{12},\quad \text{for all }x\in\mathbb{R},
    \end{equation*}
    where $\mu(x,\lda)$ is the unique solution of the above normalised Riemann-Hilbert problem with $\mu(x,\lda)=\1+\frac{\mu_1(x)}{\lda}+O\left(\frac{1}{\lda^2}\right)$ as $\lda\to\infty$.
\end{definition}
The following fundamental result yields the relation between the direct and inverse parts of the ISM at some fixed time $t=0$.
\begin{theoremm}\cite{deift2011long}
The maps $S$ and $P$ are continuous and inverse to each other.
\end{theoremm}

\subsection{Time evolution}
The effectiveness of the method now comes from the compatibility of the above construction with the time evolution: as the initial data $u$ evolves in time according to NLS, the scattering data evolves in time linearly. Specifically, we have the following.
Given a generic potential $u(x)\in \mathcal{S}(\mathbb{R})$, the solution $u(x,t)$ of the NLS equation with $u(x,t=0)=u(x)$ is generic, belongs to $\mathcal{S}(\mathbb{R})$ and the scattering data associated to $u(x,t)$ or $Q(x,t)$ for all $t\geq 0$ are given by
\begin{equation}\label{time_evol_scattdata2}
     r(t,\lda)=r(\lda)e^{2i\lda^2t}\text{ and }c(t,\lda_k)=c(\lda_k)e^{2i\lda_k^2t}.
\end{equation}
Therefore the ISM can be summarized as follows: apply the map $S$ to $u(x)\in \mathcal{S}(\mathbb{R})$, evolve the scattering data in time using formulae \eqref{time_evol_scattdata2}, and then apply the map $P$ to obtain the solution $u(x,t)$ at time $t>0$ of the NLS equation \eqref{bulk_NLS_eqs} with $u(x,t=0)=u(x)$.

\subsection{Multisoliton solutions \label{sec:ms}}
The scattering coefficient $a(\lda)$ associated with the pure multi-soliton solutions (i.e. $r(\lda)=0$ identically) reads
\begin{equation*}
    a(\lda)=\prod_{k=1}^n\frac{\lda-\lda_k}{\lda-\lda_k^*}\,.
\end{equation*}
where $\lda_1,\dots,\lda_n\in \mathbb{C}^+$ are the zeros of $a(\lda)$ in the upper half-plane.
In this case, the map $P$ is known explicitly and one gets the pure $n$-soliton as (see e.g. \cite{ablowitz2004discrete, faddeev2007hamiltonian})
    \begin{equation}
    \label{expression_u}
        u(x,t)=2i\frac{\det P_b}{\det P},
    \end{equation}
where    $P_b=\begin{pmatrix}
    0 & \textbf{c}\\
    \mathbf{1} & P^T
    \end{pmatrix}$, with $\mathbf{1}=(1,\dots,1)^T$, \small $\textbf{c}=-\left(c^*(\lda_1)e^{-2i(\lda_1^*x+2\lda_1^{*2}t)},\dots,c^*(\lda_n)e^{-2i(\lda_n^*x+2\lda_n^{*2}t)}\right),$ \normalsize 
    and
    \begin{equation*}
        P=\left(p_{m,j}\right)_{1\leq m,j\leq n},\quad
        p_{m,j}=\delta_{m,j}+c^*(\lda_m)e^{-2i(\lda_m^*x+2\lda_m^{*2}t)}\sum_{k=1}^n\frac{c(\lda_k)e^{2i(\lda_kx+2\lda_k^2t)}}{(\lda_m^*-\lda_k)(\lda_k-\lda_j^*)}\ .
    \end{equation*}

\section{B\"acklund transformation approach to integrable BCs}\label{BT_review}
	
To apply the nonlinear mirror image method to the time dependent conditions considered in this paper, we need to construct a B\"acklund matrix whose the value at $x=0$ equals $K(t,\lda)$ with $K(t,\lda)$ given by \eqref{formK} and depends on time. This poses a challenge if one tries to use the traditional approach. Indeed, usually, the B\"acklund transformation is computed at $t=0$ using the differential equation it should satisfy and imposing the boundary value at $x=0$, and then evolve it in time in a compatible manner. This works well if the boundary value at $x=0$ is indeed time-independent (as in the Robin case) but appears to be a difficult task in our time-dependent case. 

Therefore, we will employ a different strategy to overcome this problem. The idea is to exploit the fact that we can fix the boundary value of the B\"acklund transformation as $x\to\infty$, which is time-independent even in our case, as opposed to the value at $x=0$. This discussion suggests that we need to first review the properties of B\"acklund transformations in detail in a way that is tailored for our needs. This is what we do in the first subsection before moving on to revisiting the Robin case to show how our new strategy brings completely equivalent results (as it should).

\subsection{Review of B\"acklund transformations}

\subsubsection{Transformation of the potential}
In what follows $Q(x)$ denotes a $2\times 2$ off-diagonal matrix of the form $Q(x)=\begin{pmatrix}
    0&u(x)\\ -u^*(x)&0
\end{pmatrix}$ with $u(x)\in \mathcal{S}(\mathbb{R})$. We may have the case $u(x)\in \mathcal{S}(\mathbb{R}^\pm)$.
We start with the following lemma which is at the basis of our construction of the required B\"acklund transformation.
\begin{lemma}
	\label{propBT}
	Let $u(x)\in \mathcal{S}(\mathbb{R})$ and suppose that the $2\times 2$ matrix $P(x)$ satisfies the differential equation
	\begin{equation}
    	\label{eqP}
    	P_{x}=\left[\frac{i\rho}{2}\sigma_3-Q+i\sigma_3 P\sigma_3,P\sigma_3\right]\sigma_3\,,
	\end{equation}
	with $\rho\in \mathbb{R}$ and
	\begin{equation}\label{Value_of_P}
	    \lim_{x\to+\infty}P(x)=\frac{i\gamma_+}{2}\1,\quad\gamma_+\in \mathbb{R}\backslash\{0\}.
	\end{equation}
	Then, $P(x)$ has the following properties:
	\begin{enumerate}
		\item[\emph{(a)}] diagonal asymptotic at $-\infty$,
		\begin{equation*}
		    \lim_{x\to-\infty}P(x)=\frac{i\gamma_-}{2} \1\,, \text{ with }\gamma_-^2=\gamma_+^2\,.
		\end{equation*}
		\item[\emph{(b)}] $[\sigma_3,P(x)\sigma_3]\in\cS(\RR)$.
	\end{enumerate}
\end{lemma}

This is of course a standard result of the dressing method. However, for the reader's convenience, we found it useful to give a self-contained proof presented in Appendix \ref{proof_lemma}. It also provides the details we need concerning the various cases $\gamma_-=\pm\gamma_+$ that we could not find in one place in the literature. They are summarised as follows:
	\begin{enumerate}
		\item[a.] If $\gamma_+<0$ and $\lda_+=-\frac{\rho+i\gamma_+}{2}$ is not a simple zero of $a(\lda)$ then 
		$\begin{cases}
				\gamma_- = \gamma_+\,,& \text{if }\mu_2=0~\text{in}~\eqref{constant_mu}\,,\\
				\gamma_- = -\gamma_+\,,&\text{if }\mu_2\neq 0~\text{in}~\eqref{constant_mu}\,.
			\end{cases}$
		
		The first case $\mu_2=0$ is rarely mentioned in the literature since it is rather ``useless'' from the point of the view of the dressing method: it does not create a new zero for $a(\lda)$, see formula \eqref{BT1_eq_a_atilde} below. However, it does allow for the case $\gamma_- = \gamma_+$ when $\gamma_+<0$ and $\lda_+=-\frac{\rho+i\gamma_+}{2}$ is not a simple zero of $a(\lda)$, a case that cannot be overlooked in construction of the mirror image approach for the time-dependent BCs in Section \ref{BT2_section}.
		
		\item[b.] If $\gamma_+<0$ and $\lda_+$ is a simple zero of $a(\lda)$ then $\gamma_- = -\gamma_+$.
		
		\item[c.] If $\gamma_+>0$ and $\lda_+$ is not a simple zero of $a^*(\lda)$ then $\gamma_- = \gamma_+$. 
		
		\item[d.] If $\gamma_+>0$ and $\lda_+$ is a simple zero of $a^*(\lda)$ then $\gamma_- = -\gamma_+$.
	\end{enumerate}
	Case a. shows a small subtlety related to our approach of fixing $P(x)$ by its limit at $\infty$. The freedom in $\mu_2$ indicates that $P(x)$ is not uniquely determined when $\gamma_+<0$ and $u(x)$ is such that $\lda_+$ is not a simple zero of $a(\lda)$. In general, one would also need to specify whether $\mu_2=0$ or not. As we explained, if the goal was to create a soliton on a given background solution $u(x)$, one would naturally choose $\mu_2\neq 0$. This freedom will not be a problem for the application of the B\"acklund transformation to the half-line problem. The additional symmetry coming from the folding of $u(x)$ will fix uniquely the structure of $a(\lda)$ relative to whether $\gamma_- = \gamma_+$ or $\gamma_- = -\gamma_+$. With this in mind, we proceed with the fact that $P(x)$ can be constructed uniquely as in the above proposition (fixing $\mu_2$ as required if we are in case a.).   
	
Given a potential $Q(x)$ as above, we get $P(x)$ as the solution of the differential equation \eqref{eqP}-\eqref{Value_of_P}, and we define the following B\"acklund matrix
\begin{equation}
    \label{choiceL}
    L(x,\lda)=\left(\lda+\frac{\rho}{2}\right)\sigma_3+P(x)\,,~~\rho\in\RR\,.
\end{equation}
The differential equation satisfied by $P(x)$ is equivalent to $L(x,\lda)$ solving the familiar gauge transformation equation 
\begin{equation}
\label{new_eqL}
        L_x(x,\lda)=\widetilde{U}(x,\lda)L(x,\lda)-L(x,\lda)U(x,\lda)\,,
\end{equation}
where $\widetilde{U}(x,\lda)=-i\lda\sigma_3+\widetilde{Q}(x)$ with $\widetilde{Q}(x)$ given by 
		\begin{equation}
		\label{def_Qtilde}
		  \widetilde{Q}(x)=-Q(x)+i[\sigma_3,P(x)\sigma_3]\,.
		\end{equation} 
		In turn, this ensures that 
if $\Psi(x,\lda)$ is a solution of \eqref{LP1} at $t=0$, and we define 
\begin{equation}\label{BT_on_eigenfunctions}
    \widetilde{\Psi}(x,\lda):=L(x,\lda)\Psi(x,\lda)\,,
\end{equation}
then $\widetilde{\Psi}(x,\lda)$ solves $\widetilde{\Psi}_x(x,t)=\widetilde{U}(x,\lda)\widetilde{\Psi}(x,\lda)$. 
	The new potential $\widetilde{Q}(x)$ given by \eqref{def_Qtilde} belongs to $\cS(\RR)$ as a consequence of Lemma \ref{propBT}. 
\begin{definition}\label{defBT}
The map
	$$\begin{array}{cccc}
L_{\rho,\gamma_+}: & \mathcal{S}(\mathbb{R}) & \longrightarrow & \mathcal{S}(\mathbb{R})\\
    &  Q & \longmapsto  & \widetilde{Q}=L_{\rho,\gamma_+}[Q],
  \end{array}$$ 
is called the B\"acklund transformation (BT) of $Q(x)$ with respect to $(\rho,\gamma_+)$. We will use the same terminology and notation at the level of the entries $\widetilde{u}(x)$ and $u(x)$.
\end{definition}
The content of Lemma \ref{propBT}, part $(b)$ can be restricted to the half-line $\mathbb{R}^+$ with $u(x)\in \mathcal{S}(\mathbb{R}^+)$. This gives a matrix $L(x,\lda)$ as in \eqref{choiceL} and defined on $\mathbb{R}^+$. We denote the corresponding map as $L_{\rho,\gamma_+}^+:Q\mapsto\widetilde{Q}=L_{\rho,\gamma_+}^+[Q]$. Similarly, we can restrict to $\mathbb{R}^-$, with $u(x)\in\mathcal{S}(\mathbb{R}^-)$ but with the understanding that we could fix $P(x)$ at $-\infty$, \ie $P(x)\to \frac{i\gamma}{2}\1$ as $x\to -\infty$, $\gamma\in \mathbb{R}\backslash\{0\}$. The corresponding B\"acklund transformation of $Q(x)$ ($x<0$) with respect to $(\rho,\gamma)$ will be denoted by the map $L_{\rho,\gamma}^-:Q\mapsto \widetilde{Q}=L_{\rho,\gamma}^-[Q]$. 
\begin{lemma}\cite{deift2011long}\hfill
\label{Bijection_of_BT1}
\begin{enumerate}
    \item If $Q(x)\in \mathcal{S}(\mathbb{R})$ then $\mathcal{R}L_{-\rho,\gamma_-}\mathcal{R}L_{\rho,\gamma_+}[Q]=Q,$ 
    \item If $Q(x)\in \mathcal{S}(\mathbb{R}^\pm)$ then         $\mathcal{R}L_{\rho,\gamma_\pm}^\pm[Q](x)\in \mathcal{S}(\mathbb{R}^\mp)$, $\mathcal{R}L_{-\rho,\gamma_-}^-\mathcal{R}L_{\rho,\gamma_+}^+[Q]=Q,$ and $\mathcal{R}L_{-\rho,\gamma_+}^+\mathcal{R}L_{\rho,\gamma_-}^-[Q]=Q,$
\end{enumerate}
where $\mathcal{R}Q(x)\equiv -Q(-x)$.
\end{lemma}
\prf
The strategy of this proof is similar to the one given in \cite{deift2011long}. So, we highlight the main differences. Assume $Q(x)\in \mathcal{S}(\mathbb{R})$ and let $P(x)$ be the solution of \eqref{eqP}-\eqref{Value_of_P}. Define
\begin{equation*}
	Q_2(x)=-\widetilde{Q}(-x)=\mathcal{R}L_{\rho,\gamma_+}[Q](x),\quad P_2(x)=\sigma_3P(-x)\sigma_3.
\end{equation*}
We have $Q(-x) =-\widetilde{Q}(-x)+i[\sigma_3,P(-x)\sigma_3]
	 =-\left(-Q_2(x)+i[\sigma_3,P_2(x)\sigma_3]\right).$
The matrix $P(x)$ admits the symmetry $P(x)=-\sigma_3 P^\dagger(x)\sigma_3$. A direct calculation shows that
\begin{align*}
	\left(P_2(x)\right)_x 
	&=-\frac{i\rho}{2}\left[\sigma_3,P_2(x)\right]+\left(-Q_2(x)+i[\sigma_3,P_2(x)\sigma_3]\right) P_2(x) -P_2(x)Q_2(x)
\end{align*}
Taking into consideration the fact that $\displaystyle \lim_{x\to+\infty}P_2(x)=\lim_{x\to-\infty}P(x)=\frac{i\gamma_-}{2}\1$, we conclude that $\widetilde{Q}_2(x)=-Q_2(x)+i[\sigma_3,P_2(x)\sigma_3]=-Q(-x)$ which means $\mathcal{R}L_{-\rho,\gamma_-}[Q_2](x)=Q(x)$ and proves the first point of the Lemma. The second point is proven similarly.
\finprf 
It follows from this lemma that the map $ L_{\rho,\gamma_+}:\mathcal{S}(\mathbb{R})\to \mathcal{S}(\mathbb{R})$
is a bijection. The same conclusion can be drawn for $L_{\rho,\gamma_\pm}^\pm:\mathcal{S}(\mathbb{R}^\pm)\to \mathcal{S}(\mathbb{R}^\pm)$. Note that if $Q(x)\in \mathcal{S}(\mathbb{R})$, then $\left(L_{\rho,\gamma_+}[Q](x)\right)|_{\mathbb{R}^\pm}=L^{\pm}_{\rho,\gamma_\pm}\left[Q|_{\mathbb{R}^{\pm}}\right](x)$.

\subsubsection{Transformation of the scattering data}

Here we review the relation between the scattering data associated to $u(x)$ and the data associated to $\widetilde{u}(x)=L_{\rho,\gamma_+}[u](x)$, the latter being designated with a tilde on the relevant objects.
\begin{lemma}

The relation between the scattering matrix $S(\lda)$ associated to $u(x)$ and the scattering matrix $\widetilde{S}(\lda)$ associated to $\widetilde{u}(x)$ reads
\begin{equation}
	\label{relationS}
	\widetilde{S}(\lda)=\left((2\lda+\rho)\sigma_3+i\gamma_-\1\right)\,S(\lda)\,\left((2\lda+\rho)\sigma_3+i\gamma_+\1\right)^{-1}\,.
\end{equation}
Explicitly, for the coefficients, one gets
\begin{equation}\label{BT1_eq_b_btilde}
	\widetilde{b}(\lda)=-\frac{2\lda+\rho-i\gamma_-}{2\lda+\rho+i\gamma_+}b(\lda),\quad \lda\in \mathbb{R},
\end{equation} 
\begin{equation}\label{BT1_eq_a_atilde}
	\widetilde{a}(\lda)=\frac{2\lda+\rho-i\gamma_-}{2\lda+\rho-i\gamma_+}a(\lda),\quad \lda\in \mathbb{C}^+\backslash\left\{\frac{-\rho+i\gamma_+}{2}\right\}.
\end{equation}
In the case $\gamma_-=\gamma_+$, $a(\lda)$ and $\widetilde{a}(\lda)$ have the same zeros $\lda_1,\dots,\lda_n$. The norming constants $\gamma(\lda_1),\dots,\gamma(\lda_n)$ associated to $u(x)$, and $\widetilde{\gamma}(\lda_1),\dots,\widetilde{\gamma}(\lda_n)$ associated to
$\widetilde{u}(x)$ are related by
\begin{equation}\label{normingfin}
	\widetilde{\gamma}(\lda_k)=-\frac{2\lda_k+\rho-i\gamma_+}{2\lda_k+\rho+i\gamma_-}\gamma(\lda_k).
\end{equation}
\end{lemma}
\prf See for instance the proof of \cite[Proposition 4.10]{deift2011long}. \finprf

\begin{prop} 
	\label{propP}	
	Let $u(x)\in \mathcal{S}(\mathbb{R})$ and $P(x)$ be a solution of \eqref{eqP}-\eqref{Value_of_P}. The properties of $P(x)$ imply that we have the following useful explicit representation of $L$ in terms of $u$ and $\widetilde{u}=L_{\rho,\gamma_+}[u]$, 
	\begin{equation}
		\label{formL}
		L(x,\lda)=\left(\lda+\frac{\rho}{2}\right)\sigma_3+P(x)\,,~~P(x)=\frac{i}{2}\begin{pmatrix}
			\epsilon(x) \sqrt{\gamma_+^2-|\widetilde{u}+u|^2}& \widetilde{u}+u\\
			-(\widetilde{u}+u)^* & \epsilon(x) \sqrt{\gamma_+^2-|\widetilde{u}+u|^2}
		\end{pmatrix}\,,
	\end{equation}
	where $\epsilon(x)$ is a sign function completely determined by $\gamma_+$ and $u$. By construction, we have
	$$\gamma_+^2-|\widetilde{u}(x)+u(x)|^2\ge 0\,,~~\forall x\in\RR\,.$$
	Finally,
	\begin{equation}
		\label{inverseL}
		L^{-1}(x,\lda)=\sigma_3\frac{\left(\lda+\frac{\rho}{2}\right)\sigma_3-P(x)}{\left(\lda+\frac{\rho}{2}\right)^2+\frac{\gamma_+^2}{4}}\sigma_3.
	\end{equation}
\end{prop}
\prf
This result was given e.g. in \cite{caudrelier2008systematic} but we give here more details, especially on the function $\epsilon(x)$ which takes value $\pm 1$. The starting point is the construction of $P(x)$ as in Appendix \ref{proof_lemma}, see \eqref{deift_lemma_P_form}, which gives
$$P(x)=\begin{pmatrix}
    C(x) & D(x)\\
    D^*(x) & C(x)
\end{pmatrix}\,,~~C(x)=\frac{i\gamma_+}{2}\frac{|\xi_1(x)|^2-|\xi_2(x)|^2}{|\xi_1(x)|^2+|\xi_2(x)|^2}\,,~~D(x)=-\frac{i\gamma_+}{2}\frac{\xi_2(x)^*\xi_1(x)}{|\xi_1(x)|^2+|\xi_2(x)|^2}\,.$$
A direct calculation then gives that $\det P(x)=-\frac{\gamma_+^2}{4}\1$ so that $C(x)^2=-\frac{\gamma_+^2}{4}+|D(x)|^2$. Since $C^*(x)=-C(x)$, we deduce $|C(x)|^2=\frac{\gamma_+^2}{4}-|D(x)|^2\ge 0$.
Next, formula \eqref{def_Qtilde} gives $D(x)=\frac{i}{2}(u(x)+\tilde{u}(x))$. Hence, 
$$\gamma_+^2-|u(x)+\tilde{u}(x)|^2\ge 0\,.$$
Combining everything, we have 
$C(x)=\frac{i\epsilon(x)}{2}\sqrt{\gamma_+^2-|u(x)+\tilde{u}(x)|^2}\,,$
where $\epsilon(x)^2=1$ gives the sign in front of the square root. We can determine its value by comparing the two expressions for $C(x)$, yielding
$$\epsilon(x)|\gamma_+|\sqrt{1-\frac{1}{\gamma_+^2}|u(x)+\tilde{u}(x)|^2}=\gamma_+\frac{|\xi_1(x)|^2-|\xi_2(x)|^2}{|\xi_1(x)|^2+|\xi_2(x)|^2}\,.$$
The sign of the expression on the LHS is of course $\epsilon(x)$ by construction. The sign of the expression on the RHS is the product of the sign of $\gamma_+$ and that of $|\xi_1(x)|^2-|\xi_2(x)|^2$. The latter is completely determined by $u(x)$.
Finally, \eqref{inverseL} is a consequence of the fact that $(P(x)\sigma_3)^2=-\frac{\gamma_+^2}{4}\1$ as is checked directly.
\finprf
\subsubsection{Time evolution}
The construction of a B\"acklund transformation is useful if it is compatible with the time evolution of the PDE of interest. 
 Consider $Q(x,t)\in \mathcal{S}(\mathbb{R})$ subject to $U_t-V_x+[U,V]=0$. For each $t\geq 0$, construct $P(x,t)$ as the solution of \eqref{eqP}-\eqref{Value_of_P}, and hence also the corresponding $L(x,t,\lda)$ which then satisfies \eqref{new_eqL}. In line with Definition \ref{defBT}, define then the new potential $\widetilde{Q}(x,t)=-Q(x,t)+i[\sigma_3,P(x,t)\sigma_3]$, for each $t\geq 0$. Call it the B\"acklund transformation of $Q(x,t)$ via $(\rho,\gamma_+)$ and write $\widetilde{Q}(x,t)=L_{\rho,\gamma_+}[Q](x,t)$. Then, the following well known result shows that the new potential also satisfies NLS if and only if $L$ satisfies the $t$-part of the gauge transformation equation. Specifically, we have
\begin{lemma}\label{time_evol} The following equivalence holds:
\begin{equation} 
    \widetilde{U}_t-\widetilde{V}_x+[\widetilde{U},\widetilde{V}]=0 \Longleftrightarrow
        L_t(x,t,\lda)=\widetilde{V}(x,t,\lda)L(x,t,\lda)-L(x,t,\lda)V(x,t,\lda)
\end{equation}

where $\widetilde{V}$ is given by replacing $Q$ by $\widetilde{Q}$ in \eqref{cont_pair2}.    
\end{lemma}
\prf Indeed, we start by proving the implication from the left to the right by assuming that $L$ satisfies the $t$-part of the gauge transformation equation.
Since $L$ also satisfies \eqref{new_eqL}, the compatibility $L_{xt}
    =L_{tx}$ yields 
\begin{equation*}
    \left(\widetilde{U}_t-\widetilde{V}_x+[\widetilde{U},\widetilde{V}]\right)L=L\left(U_t-V_x+[U,V]\right)=0\,. 
\end{equation*}
Hence, $\widetilde{U}_t-\widetilde{V}_x+[\widetilde{U},\widetilde{V}]=0$.
Conversely, assume that $\widetilde{U}_t-\widetilde{V}_x+[\widetilde{U},\widetilde{V}]=0$. Set $\Delta=L_t-\widetilde{V}L+LV$. An explicit calculation gives 
\begin{equation}
    \label{Delta_def}
    \Delta=P_t-\rho\left(\frac{i\rho}{2}[\sigma_3,P]-\widetilde{Q}P-PQ\right)+i\left(\widetilde{Q}_x\sigma_3+\widetilde{Q}^2\sigma_3\right)P-iP\left(Q_x\sigma_3+Q^2\sigma_3\right)\,,
\end{equation}
which shows that $\Delta$ does not depend on $\lda$. Now, $0=\left(\widetilde{U}_t-\widetilde{V}_x+[\widetilde{U},\widetilde{V}]\right)L= i\lda[\sigma_3,\Delta]+\Delta_x-\widetilde{Q}\Delta+Q\Delta$.
Since $\Delta$ does not depend on $\lda$, the last equation gives us $[\sigma_3,\Delta]=0$ and $\Delta_x=\widetilde{Q}\Delta-Q\Delta$. The former equation means that $\Delta$ is diagonal. The latter, as a consequence, implies that $\Delta$ is constant with respect to $x$ since the term on the right-hand side is off-diagonal. Thus, we can evaluate the constant value of $\Delta$ using \eqref{Delta_def} as $x\to\infty$. Since $Q(x)\in\mathcal{S}(\mathbb{R})$ and $\displaystyle\lim_{x\to+\infty} P(x,t,\lda)=\frac{i\gamma_+}{2}\1$, we find $\Delta=0$ as desired.
\finprf

\subsection{Revisiting the case of Robin boundary  conditions}
Consider the initial-boundary value problem for the NLS on a half line with Robin BCs
\begin{equation}\label{NLS_IBVP_Robin}
    \begin{cases}
        iu_t+u_{xx}+2|u|^2u=0 \quad\text{for } x\geq0,\,~~t\ge 0,\\
         u(x,0)=u_0(x)\in\cS(\RR^+) \quad \text{(initial condition)},\\
        u_x(0,t)+\theta u(0,t)=0\,,~~t\ge 0,\quad\theta\in\mathbb{R}\backslash\{0\},~~\text{(Robin BCs)}.
    \end{cases}
\end{equation}
As mentioned in the introduction, the use of B\"acklund transformation to map the above initial-boundary value problem to an initial value problem has been successfully studied, see for example \cite{bikbaev1991initial,deift2011long}. In what follows, we connect this method with the Sklyanin's approach to integrable BCs \cite{sklyanin1987boundary} and then show how the matrix $L(x,\lda)$ that we constructed in Lemma \ref{propBT} induces this map. From Sklyanin's work \cite{sklyanin1987boundary}, one knows that Robin BCs are equivalent to 
\begin{equation}
\label{VK_Robin}    
V(0,t,-\lda)K(\lda)-K(\lda)V(0,t,\lda)=0\,,
\end{equation}
where, in this subsection, the boundary $K$ matrix is the one associated to Robin BCs, \ie
\begin{equation}\label{eq:Krobin}
    K(\lda)=\lda\sigma_3+\frac{i\theta}{2}\1\,.
\end{equation}
To map the above initial-boundary value problem on the half-line to an initial problem value problem using the B\"acklund transformation approach, it suffices to construct a B\"acklund matrix $L(x,t,\lda)$ such that: 
\begin{enumerate}
    \item $L(x,t,\lda)$ has a diagonal time-independent value at $x=0$ given by 
    \begin{equation}
    \label{Kat0}
    L(0,t,\lda)=K(\lda)\,,
\end{equation}
\item  if the transformed Lax pair under $L$ is denoted by $(\widetilde{U},\widetilde{V})$, we have
\begin{equation}
    \label{symmetryUV}
    \widetilde{U}(-x,t,-\lda)=-U(x,t,\lda)\,,~~\widetilde{V}(-x,t,-\lda)=V(x,t,\lda)\,.
\end{equation}
\end{enumerate}
Indeed, using the time evolution equation for $L$, \ie 
\begin{equation}
    \label{Lt}
    L_t(x,t,\lda)=\widetilde{V}(x,t,\lda)L(x,t,\lda)-L(x,t,\lda)V(x,t,\lda)
\end{equation} 
and evaluating at $x=0$ with \eqref{Kat0} and \eqref{symmetryUV}, one obtains \eqref{VK_Robin} which is equivalent to Robin BCs. 
As already said, since the matrix $K(\lda)$ on the right of Eq. \eqref{Kat0} is time-independent, the construction of $L$ can be done at time $t=0$ using equation $L_x=\widetilde{U}L-LU$ and fixing its value at $x=0$ as in \eqref{Kat0}. This leads to a time-independent diagonal value of $L$ as $x\to\infty$.
The following result is a well-known fact.
\begin{lemma}
	Let $Q(x)$ be an element of $\mathcal{S}(\mathbb{R})$ and $\Psi(x,\lda)$ a $2\times 2$ invertible solution of \eqref{LP1} at time $t=0$. Then 
	    \begin{equation}\label{S_limit}
		S(\lda)=\lim_{x\to\infty}e^{-i\lda\sigma_3x}\Psi(-x,\lda)\Psi^{-1}(x,\lda)e^{-i\lda\sigma_3x}.
	\end{equation}

\end{lemma}
\prf A proof can be found in \cite[Lemma 4.27]{deift2011long}. \finprf

To apply the above strategy in order to solve the initial-boundary value problem \eqref{NLS_IBVP_Robin} but with the B\"acklund matrix $L(x,\lda)$ as we constructed it in Lemma \ref{propBT} at time $t=0$, we need the following result. It relates our construction using the boundary condition for $L(x,\lda)$ as $x\to\infty$ infinity to the traditional construction fixing $L(0,\lda)$.  
\begin{lemma}\label{SamelimiteBT1}
	Let $u(x)\in\cS(\RR)$ be such that its scattering coefficient $a(\lda)$ has $n$ simple zeros $\lda_1,\dots,\lda_n$ in $\CC^+$. Let $L(x,\lda)$ be given as in \eqref{choiceL} where $P(x)$ solves \eqref{eqP}-\eqref{Value_of_P}. If  $\widetilde{u}(x)=L_{\rho,\gamma_+}[u](x)$ is such that 
	\begin{equation} \label{beta_symmetric}
	    \widetilde{u}(x)=-u(-x)   
	\end{equation}
	holds, then
	\begin{equation}
			\gamma_-=\gamma_+\equiv \gamma\,,~~ \rho=0\,,~~ \text{ and }~~		L(0,\lda)=\lda\sigma_3+\frac{i(-1)^n\gamma}{2}\1\,.
	\end{equation}
\end{lemma}
\prf
If \eqref{beta_symmetric} is satisfied then $\widetilde{U}(-x,-\lda)=-U(x,\lambda)$. Jost solutions are then related by
\begin{equation}\label{BT_JostSols}
	\widetilde{\Psi}_\pm(-x,-\lda)=\Psi_\mp(x,\lambda).
\end{equation}
From Lemma \ref{propBT}, Eqs. \eqref{BT_on_eigenfunctions} and (\ref{BT_JostSols}), one then deduces 
\begin{equation*}\label{RBCs_JostSolBackTransf}
	\Psi_\pm(x,\lambda)=L(x,\lambda)^{-1}\Psi_\mp(-x,-\lda)\left(\left(\lda+\frac{\rho}{2}\right)\sigma_3+\frac{i\gamma_\pm}{2}\1\right),\quad \lda\in\mathbb{R}.
\end{equation*}
It follows that $\Psi_+(x,\lambda)  = L(x,\lambda)^{-1}L(-x,-\lambda)^{-1}\Psi_+(x,\lda)\left(\left(-\lda+\frac{\rho}{2}\right)\sigma_3+\frac{i\gamma_-}{2}\1\right)\left(\left(\lda+\frac{\rho}{2}\right)\sigma_3+\frac{i\gamma_+}{2}\1\right)$, which we rewrite as
\begin{equation}
    \Psi_+(x,\lda)^{-1}L(-x,-\lambda)L(x,\lambda)\Psi_+(x,\lambda)  = \left(\left(-\lda+\frac{\rho}{2}\right)\sigma_3+\frac{i\gamma_-}{2}\1\right)\left(\left(\lda+\frac{\rho}{2}\right)\sigma_3+\frac{i\gamma_+}{2}\1\right)\,.
\end{equation}
In fact, the latter relation is a consequence of the more general property that $\Psi(x,\lda)^{-1}L(-x,-\lambda)L(x,\lambda)\Psi(x,\lambda)$ is independent of $x$ for any fundamental solution of $\Psi_x=U\Psi$, as a direct calculation shows. In particular, we must have
\begin{equation}
  \left(\left(-\lda+\frac{\rho}{2}\right)\sigma_3+\frac{i\gamma_-}{2}\1\right)\left(\left(\lda+\frac{\rho}{2}\right)\sigma_3+\frac{i\gamma_+}{2}\1\right)=    \Psi_+(0,\lda)^{-1}L(0,-\lambda)L(0,\lambda)\Psi_+(0,\lambda)\,.
\end{equation}
From \eqref{def_Qtilde}, we have that $P(0)$ is diagonal when $\widetilde{Q}(x)+Q(x)=0$ and from the properties of $P(x)$ established in Lemma \ref{propBT}, we see that $P(0)=\frac{i\nu}{2}\1$ with $\nu^2=\gamma_+^2$. Therefore, we obtain the condition
\begin{equation}
2i\lambda(\gamma_--\gamma_+)\sigma_3+i\rho(\gamma_-+\gamma_+)\sigma_3-\gamma_-\gamma_+\1=    \Psi_+(0,\lda)^{-1}(2i\rho \nu\sigma_3)\Psi_+(0,\lambda)-\nu^2\1\,.
\end{equation}
Taking the trace and using $\nu^2=\gamma_+^2$ we deduce $\gamma_+=\gamma_-\equiv \gamma$. Another consequence of \eqref{BT_JostSols} is that $\widetilde{S}(\lambda)=S^{-1}(-\lambda)$ so in particular $\widetilde{b}(\lambda)=-b(-\lambda)$ and $\widetilde{a}(\lda)=a^*(-\lda)$. Combined with \eqref{BT1_eq_b_btilde} and \eqref{BT1_eq_a_atilde}, we obtain $b(-\lda)=\frac{2\lda+\rho-i\gamma}{2\lda+\rho+i\gamma}b(\lda)$ and $a^*(-\lda^*)=a(\lda)$. The symmetry for $b(\lda)$ is consistent iff $(\rho-i\gamma)^2=(\rho+i\gamma)^2$. Since $\gamma\neq 0$ this implies $\rho=0$. For any fundamental solution $\Psi(x,\lda)$ relation \eqref{BT_JostSols} takes the form $\widetilde{\Psi}(-x,-\lda)=\Psi(x,\lda)M(\lda)$ for some matrix $M(\lda)$. Evaluating the latter equation at $x=0=\lda$ and using \eqref{BT_on_eigenfunctions}, one obtains $M(0)=\frac{i\nu}{2}$. Hence, using \eqref{BT_on_eigenfunctions}, one gets $\displaystyle S(0)=\lim_{x\to\infty}\Psi(-x,0)\Psi(x,0)^{-1}=\frac{\gamma}{\nu}$. Evaluating relation \eqref{relationS} at $\lda=0$ and using $\widetilde{S}(\lda)=S(-\lda)^{-1}$, one obtains $S(0)=a(0)\1$ which means $a(0)=\frac{\gamma}{\nu}$. The scattering coefficient $a(\lda)$ has the following explicit form \cite{ablowitz2004discrete}
\begin{equation*}\label{trace_formu1}
	a(\lda)=\prod_{k=1}^{n}\frac{\lda-\lda_k}{\lda-\lda^*_k}\exp\left(\frac{1}{2\pi i}\int_{\mathbb{R}}\frac{\log\left(1-|b(z)|^2\right)}{z-\lda}\text{d}z\right), \qquad \text{Im}(\lda)>0.
\end{equation*} 
Computing $\displaystyle \lim_{\underset{\text{Im}\lda>0}{\lda\to 0}}a(\lda)$ from the above formula and using $|b(\lda)|=|b(-\lda)|$, one obtains $\displaystyle a(0)=\prod_{k=1}^{n}\frac{\lda_k}{\lda^*_k}$. Combined this with $a(\lda)=a^*(-\lda^*)$, one deduces $a(0)=(-1)^n$. Hence, one has $\nu=(-1)^n\gamma$. This completes the proof.
\finprf

This lemma shows that the matrix transformation $L(x,\lda)$ we constructed in Lemma \ref{propBT} has a diagonal value at $x=0$ which is equal to $K(\lda)$ upon setting $\gamma_+=\gamma_-\equiv\gamma$ to $(-1)^n\theta$. Therefore, in the case of Robin BCs, fixing the boundary value of the B\"acklund transformation at $x=0$ or as $x\to\infty$ does not make any difference. As Bikbaev and Tarasov noticed in \cite{bikbaev1991initial}, relation \eqref{beta_symmetric} can be reformulated in terms of scattering data associated to $u(x)$ as follows:
\begin{prop}
	\label{symmetrydata}
Let $u(x)\in\cS(\RR)$ be such that its scattering coefficient $a(\lda)$ has $n$ simple zeros $\lda_1,\dots,\lda_n$ in $\CC^+$. Let $L(x,\lda)$ be given as in \eqref{choiceL} where $P(x)$ solves \eqref{eqP}-\eqref{Value_of_P}. Then $u(x)$ satisfies condition \eqref{beta_symmetric} if and only if the following symmetry on the scattering data holds
\begin{equation}\label{BT1_ContScatData_sym}
	a^*(-\lda^*)=a(\lda),\quad \lda\in \mathbb{C}^+,\quad b(-\lda)=\frac{2\lda-i\gamma}{2\lda+i\gamma}b(\lda),\quad \lda\in \mathbb{R},
\end{equation}
\begin{equation}\label{BT1_DisctScatData_sym}
	\lda_k\neq \pm i\gamma/2~~\text{and}~~\gamma(\lda_k)\gamma^*(-\lda_k^*)=\frac{2\lda_k+i\gamma}{2\lda_k-i\gamma}\,,\quad k=1,\dots,n.
\end{equation}
\end{prop}
\prf In view of Lemma \ref{SamelimiteBT1}, one can use the same proofs as in \cite{bikbaev1991initial,deift2011long} for instance. We do not repeat them here.\finprf

We now summarise the nonlinear mirror image strategy, as proposed in \cite{bikbaev1991initial}, to solve \eqref{NLS_IBVP_Robin}. Starting from the initial condition $u_0(x)\in \mathcal{S}(\mathbb{R^+})$ satisfying $(u_0)_x+\theta u_0=0$ at $x=0$, construct its B\"acklund transformation $\widetilde{u}_0(x)=L_{\rho,\gamma_+}^+[u_0](x)$ and introduce an extension $u^{ext}_0(x)$ to the full line by setting
\begin{equation}
	\label{ext}
	u^{ext}_0(x)=\begin{cases}
	u_0(x)\,,~~x\ge 0\,,\\
	-\widetilde{u_0}(-x)\,,~~x<0\,.
	\end{cases}
	\end{equation}
	Then: $(i)$ It follows from Lemmas \ref{Bijection_of_BT1} and \ref{SamelimiteBT1} that $u^{ext}_0(x)$ satisfies condition \eqref{beta_symmetric} and Robin boundary condition upon setting $\gamma=(-1)^n\theta$; $(ii)$ The extension $u^{ext}_0(x)$ provides valid initial data\footnote{The only technicality involves smoothness properties at $x=0$, see \cite[Proposition 4.26]{deift2011long} or Appendix D in \cite{bikbaev1991initial}} to implement the inverse scattering method on $\RR$ in order to obtain the solution $u^{ext}(x,t)$ for $t\ge 0$. The compatibility of symmetries \eqref{BT1_ContScatData_sym}-\eqref{BT1_DisctScatData_sym} with the time evolution $a(t,\lda)=a(\lda)$, $b(t,\lda)=b(\lda)e^{2i\lda^2t}$ and $\gamma(t,\lda_k)=\gamma(\lda_k)e^{2i\lda_k^2t}$ known from ISM, ensures that the condition $\widetilde{u}^{ext}(x,t)=-u^{ext}(-x,t)$ now holds for all $t\ge 0$. As a consequence, so does the boundary condition $u^{ext}_x(0,t)+\theta u^{ext}(0,t)=0$ for all $t\geq0$. The desired solution of \eqref{NLS_IBVP_Robin} is simply obtained by taking $u(x,t)=u^{ext}(x,t)|_{\RR^+}$.
	
	The B\"acklund matrix $L(x,t,\lda)$ that allowed us to map the initial-boundary value problem \eqref{NLS_IBVP_Robin} to an initial-value problem has been constructed by fixing its value as $x\to\infty$.

\section{Nonlinear mirror image method: case of the time-dependent BCs}\label{BT2_section}

\subsection{General strategy}

To tackle the time-dependent BCs given by \eqref{BCQn}, we follow the
method reviewed in detail so far.
We look for a BT induced by a matrix $B(x,t,\lda)$ such that
\begin{equation}\label{Conditions_on_B}
    B_x=\widetilde{U}B-BU\,,~~B_t=\widetilde{V}B-BV\,,
\end{equation}
with $(U,V)$ the Lax pair of NLS equation,
$(\widetilde{U},\widetilde{V})$ the transformed Lax pair under $B$, and such that 
\begin{equation}
\label{condition_B}
    B(0,t,\lda)=K(t,\lda)
\end{equation} with 
$K(t,\lda)$ given by \eqref{formK}-\eqref{formH}, when the folding symmetry
\be
\widetilde{U}(-x,t,-\lda)=-U(x,t,\lda)
\ee
is required. 
The following argument suggests how to proceed. In each step, we consider a B\"acklund transformation induced by a matrix of the form \eqref{choiceL}. 
  Consider a first B\"acklund transformation $L_1(x,\lda)$ yielding $\widehat{Q}(x)$ from $Q(x)$ \ie satisfying
 $$L_{1x}(x,\lda)=\widehat{U}(x,\lda)L_1(x,\lda)-L_1(x,\lda)U(x,\lda)\,.$$
 With $\widehat{U}(x,\lda)$ thus obtained, we can consider the transformation $L_2$ yielding $-\widehat{Q}(-x)$ from $\widehat{Q}(x)$, \ie a solution of 
 $$L_{2x}(x,\lda)=-\widehat{U}(-x,-\lda)L_2(x,\lda)-L_2(x,\lda)\widehat{U}(x,\lda)\,.$$
  It is given by $L_2(x,\lda)=\widehat{\Psi}(-x,-\lda)C_2(\lda)\widehat{\Psi}^{-1}(x,\lda)$ for some matrix $C_2(\lda)$ independent of $x$ and where $\widehat{\Psi}(x,\lda)$ is a solution of $\widehat{\Psi}_x(x,\lda)=\widehat{U}(x,\lda)\widehat{\Psi}(x,\lda)$\footnote{The exact normalisation would come from fixing a boundary condition but is not relevant for our argument.}. Finally we consider a third transformation $L_3$ producing $\widetilde{Q}(x)$ from $-\widehat{Q}(-x)$,
  $$L_{3x}(x,\lda)=\widetilde{U}(x,\lda)L_3(x,\lda)+L_3(x,\lda)\widehat{U}(-x,-\lda)\,.$$
  By construction $B(x,\lda)=L_3L_2L_1(x,\lda)$ provides a B\"acklund transformation from $Q(x)$ to $\widetilde{Q}(x)$,\ie 
  \begin{equation}\label{EqB}
    B_x(x,\lda)=\widetilde{U}(x,\lda)B(x,\lda)-B(x,\lda)U(x,\lda)\,.
\end{equation}
A direct calculation then gives
  \begin{eqnarray}
      &&\widetilde{U}(x,\lda)+U(-x,-\lda)\nonumber\\
     \qquad&=&L_1^{-1}(-x,-\lda)\left(\partial_x(L_1(-x,-\lda)L_3(x,\lda))-\left[L_1(-x,-\lda)L_3(x,\lda),\widehat{U}(-x,-\lda)\right]\right)L_3^{-1}(x,\lda)\,.\qquad\qquad
  \end{eqnarray}
  Hence the folding condition $\widetilde{U}(x,\lda)+U(-x,-\lda)=0$ is equivalent to 
  $$\partial_x(L_1(-x,-\lda)L_3(x,\lda))=\left[L_1(-x,-\lda)L_3(x,\lda),\widehat{U}(-x,-\lda)\right]\,,$$
  which in turns is equivalent to 
  $$L_1(-x,-\lda)L_3(x,\lda)=\widehat{\Psi}(-x,-\lda)C_3(\lda)\widehat{\Psi}^{-1}(-x,-\lda)\,,$$
  for some matrix $C_3(\lda)$. Using this to eliminate $L_3$ and recalling the expression of $L_2$, we obtain
  \begin{equation}
      B(x,\lda)=L_1^{-1}(-x,-\lda)\widehat{\Psi}(-x,-\lda)C_3(\lda)C_2(\lda)\widehat{\Psi}^{-1}(x,\lda)L_1(x,\lda).
  \end{equation}
  Choosing for definiteness the Jost solution $\widehat{\Psi}_+(x,\lda)$ and fixing the boundary condition for $B$ in \eqref{EqB} as 
  \begin{equation}\label{limite_of_B}
	\lim_{x\to\infty}B(x,\lda)=h(\lda) \left[(-2\lda+\rho)\sigma_3-i\gamma_-\right]\left[(2\lda+\rho)\sigma_3+i\gamma_+\right],
\end{equation}
where $h(\lda)=\frac{1}{\left(2\lda-\rho\right)^2+\gamma_+^2}$, we obtain that 
$$\lim_{x\to\infty}\widehat{\Psi}_+(-x,-\lda)C_3(\lda)C_2(\lda)\widehat{\Psi}^{-1}_+(x,\lda)=\1\,,$$
which yields $C_3(\lda)C_2(\lda)=\widehat{S}^{-1}(-\lda)$. Summarising and recalling that $\widehat{\Psi}_+(-x,-\lda)\widehat{S}^{-1}(-\lda)=\widehat{\Psi}_-(-x,-\lda)$, we have obtained that the folding symmetry is equivalent to 
\begin{equation}
      B(x,\lda)=L_1^{-1}(-x,-\lda)\widehat{\Psi}_-(-x,-\lda)\widehat{\Psi}^{-1}_+(x,\lda)L_1(x,\lda)\,.
  \end{equation}
  This suggests that of all the possible potentials $Q(x)$, those that are such that $\widehat{Q}(x)=-\widehat{Q}(-x)$ will fulfill the desired conditions. Indeed, in that special case of \eqref{BT_JostSols}, we have $\widehat{\Psi}_-(-x,-\lda)\widehat{\Psi}^{-1}_+(x,\lda)=\1$ and
  \begin{equation}
  \label{folded_B}
      B(x,\lda)=L_1^{-1}(-x,-\lda)L_1(x,\lda)\,.
  \end{equation}
It remains to check that such a $B$ can satisfy \eqref{condition_B} to ensure that it is the right candidate with the required properties. 
Using Proposition \ref{propP}, which gives
 \begin{equation}
		L_1(x,\lda)=\left(\lda+\frac{\rho}{2}\right)\sigma_3+\frac{i}{2}\begin{pmatrix}
			\epsilon_1(x) \sqrt{\gamma_+^2-|\widehat{u}+u|^2}& \widehat{u}+u\\
			-(\widehat{u}+u)^* & \epsilon_1(x) \sqrt{\gamma_+^2-|\widehat{u}+u|^2}
		\end{pmatrix}\,,
	\end{equation}
	a direct calculation shows, recalling that we work under the assumption $\widehat{u}(x)=-\widehat{u}(-x)$,
	\begin{equation}
  \label{folded_B_at_origin}
      B(0,\lda)=L_1^{-1}(0,-\lda)L_1(0,\lda)=\frac{1}{(2\lda-\rho)^2+\gamma_+^2}\left(-4\lda^2\1-4i\lda H+\gamma_+^2+ \rho^2\right)\,,
  \end{equation}
	with
\begin{equation}\label{matrixH}
    H=\begin{pmatrix}
    \epsilon_1(0)\sqrt{\gamma_+^2-|u|^2(0)}&u(0)\\
     u^*(0)& -\epsilon_1(0)\sqrt{\gamma_+^2-|u|^2(0)}
    \end{pmatrix}\,.
\end{equation}
This is the desired result, see \eqref{bd_matrix_NLS}, if we set $\rho=\beta$ and $\gamma_+^2=\alpha^2$. In view of the results of Section \ref{BT_review} about the time evolution compatibility of a B\"acklund transformation, we can conclude that our $B$ will satisfy \eqref{condition_B}. 

Summarising the discussion, we construct $B(x,\lda)$ as the product $L_1^{-1}(-x,-\lda)L_1(x,\lda)$ to realise the B\"acklund 
transformation $Q\mapsto \widetilde{Q}$ where $L_1(x,\lda)$ realises the map $Q\mapsto \widehat{Q}=L_{1\rho,\gamma_+}[Q]$ and we consider those potentials $Q(x)$ that are such that $\widehat{Q}$ is an odd function. This automatically ensures the folding condition $\widetilde{Q}(x)=-Q(-x)$. This is the analog in our case of $Q$ being ``$q$-symmetric'' in the terminology of \cite{deift2011long} who dealt with the Robin case.
In the rest of this article, we fix 
\begin{equation} \label{eq:sie}
\rho=\beta\,,~~\gamma_+=\varepsilon\alpha\,,~~\varepsilon=\pm 1\,.
\end{equation}

\subsection{Results}
Our first task is to characterise the symmetry properties of the scattering data of a potential $Q(x)$ satisfying the condition we have just discussed \ie  $\widehat{Q}=L_{1\rho,\gamma_+}[Q]$ is an odd function. Unlike the Robin case, this condition does not impose $\gamma_+=\gamma_-$ and we have to consider the two cases $\gamma_+=\pm\gamma_-$. We proceed in several steps, concentrating on the continuous data first. 
\begin{prop}\label{Relat_Q_QtildeBT2}
 Let $Q(x)\in \mathcal{S}(\mathbb{R})$ be such that $\widehat{Q}=L_{1\rho,\gamma_+}[Q]$ is an odd function. Then, its scattering data satisfies
\begin{equation}
\label{BT2relationS}
S^{-1}(-\lda)=\cB(\lda)\,S(\lda)\,\cB(-\lda)\,,~~\cB(\lda)=\begin{pmatrix}
    \frac{2\lda+\beta+i\gamma_-}{-2\lda+\beta+i\gamma_+} & 0\\
    0 & \frac{2\lda+\beta-i\gamma_-}{-2\lda+\beta-i\gamma_+}
\end{pmatrix}\,.
\end{equation}
Explicitly, if $\gamma_+=\gamma_-=\varepsilon\alpha$, we get
\begin{equation}
\label{relation_a1}
    a(-\lda)=a^*(\lda^*)\,,~~b(-\lda)=-\frac{2\lda+\beta-i\varepsilon\alpha}{2\lda+\beta+i\varepsilon\alpha}\ \frac{2\lda-\beta-i\varepsilon\alpha}{2\lda-\beta+i\varepsilon\alpha}\ b(\lda)\,.
\end{equation}
If $\gamma_+=-\gamma_-=\varepsilon\alpha$, we get
\begin{equation}
\label{relation_a2}
    a(-\lda)=\frac{2\lda+\beta-i\varepsilon\alpha}{2\lda+\beta+i\varepsilon\alpha}\ \frac{2\lda-\beta+i\varepsilon\alpha}{2\lda-\beta-i\varepsilon\alpha}\ a^*(\lda^*)\,,~~b(-\lda)=-b(\lda)\,.
\end{equation}
\end{prop}
\prf
We have the following relation between Jost solutions
\begin{equation}
\label{dressing_Psi}
    \widehat{\Psi}_\pm(x,\lda)=L_1(x,\lda)\Psi_\pm(x,\lda)\,L_\pm^{-1}(\lda)\,,~~L_\pm(\lda)=\lim_{x\to\pm\infty}L_1(x,\lda)=\left(\lda+\frac{\rho}{2}\right)\sigma_3+\frac{i\gamma_\pm}{2}\1\,.
\end{equation}
It implies in particular that $\widehat{S}(\lda)=L_-(\lda)S(\lda)L_+^{-1}(\lda)$.
Since $\widehat{U}(-x,-\lda)=-\widehat{U}(x,\lambda)$, we also have
\begin{equation}
\label{sym_psihat}
	\widehat{\Psi}_\pm(-x,-\lda)=\widehat{\Psi}_\mp(x,\lambda)\,.
\end{equation}
This implies in particular that $\widehat{S}^{-1}(-\lda)=\widehat{S}(\lda)$. Combining these two results, and recalling that $\rho=\beta$ and $\gamma_+=\varepsilon\alpha$, yields \eqref{BT2relationS} with $\cB(\lda)=L_+^{-1}(-\lda)L_-(\lda)$ as desired.
\finprf
These symmetries have consequences for the possible discrete data. We gather the result in the following Proposition.
\begin{prop}\label{prop_sym_discrete}
 Let $Q(x)\in \mathcal{S}(\mathbb{R})$ be such that $\widehat{Q}=L_{1\rho,\gamma_+}[Q]$ is an odd function. 
\begin{enumerate}
    \item If $\gamma_+=\gamma_-=\varepsilon\alpha$:

\begin{itemize}
    \item The zeros of $a(\lda)$ are composed of $p$ pairs $(\lda_k,-\lda_k^*)$, $k=1,\dots,p$ and $s$ self-symmetric zeros \cite{biondini2012nonlinear} $\lda_k=i\sigma_k\in i\RR^+$, $k=1,\dots,s$. The number $s$ of self-symmetric zeros is necessarily even. Explicitly,
    \begin{equation}
    \label{form_a1}
    a(\lda)=\displaystyle
\prod_{j=1}^p\frac{\lda-\lda_j}{\lda-\lda_j^*}\ \frac{\lda+\lda_j^*}{\lda+\lda_j}\ \prod_{k=1}^s\frac{\lda-i\sigma_k}{\lda+i\sigma_k}\ \exp\left(\frac{1}{2\pi i}\int_{\mathbb{R}}\frac{\log\left(1-|b(z)|^2\right)}{z-\lda}\text{d}z\right)\,.
\end{equation}
    
    \item The related norming constants satisfy the symmetry relation 
    \begin{equation}\label{norming-const1}
    \gamma^*(-\lda_k^*)\gamma(\lda_k)=-\frac{2\lda_k-\beta+i\varepsilon\alpha}{2\lda_k-\beta-i\varepsilon\alpha}\ \frac{2\lda_k+\beta+i\varepsilon\alpha}{2\lda_k+\beta-i\varepsilon\alpha}\,,~~k=1,\dots,2p+s\,.
\end{equation}
\end{itemize}

\item If $\gamma_+=-\gamma_-=\varepsilon\alpha$:

\begin{itemize}
    \item The zeros of $a(\lda)$ include $\frac{-\beta+i\alpha}{2}$ when $\varepsilon=1$ or $\frac{\beta+i\alpha}{2}$ when $\varepsilon=-1$, and $p$ pairs $(\lda_k,-\lda_k^*)$, $k=1,\dots,p$. There are no self-symmetric zeros. Explicitly,
    \small
    \begin{equation}
    \label{form_a2}
    a(\lda)=\begin{cases}
    \displaystyle
    \frac{2\lda+\beta-i\alpha}{2\lda+\beta+i\alpha}\ \prod_{j=1}^p\frac{\lda-\lda_j}{\lda-\lda_j^*}\ \frac{\lda+\lda_j^*}{\lda+\lda_j}\ \exp\left(\frac{1}{2\pi i}\int_{\mathbb{R}}\frac{\log\left(1-|b(z)|^2\right)}{z-\lda}\text{d}z\right)\,,~~\varepsilon=1\,,\\
    \displaystyle
    \frac{2\lda-\beta-i\alpha}{2\lda-\beta+i\alpha}\ \prod_{j=1}^p\frac{\lda-\lda_j}{\lda-\lda_j^*}\ \frac{\lda+\lda_j^*}{\lda+\lda_j}\ \exp\left(\frac{1}{2\pi i}\int_{\mathbb{R}}\frac{\log\left(1-|b(z)|^2\right)}{z-\lda}\text{d}z\right)\,,~~\varepsilon=-1\,,
    \end{cases}
\end{equation}\normalsize
    
    \item The related norming constants satisfy the symmetry relation 
    \begin{equation}
\label{no_self}
    \gamma^*(-\lda_k^*)\gamma(\lda_k)=-1\,,~~k=1,\dots,p\,.
\end{equation}
\end{itemize}

\end{enumerate}
\end{prop}
\prf
We first derive the properties of the zeros of $a(\lda)$ in each case and will provide a proof for the relation of the norming constants at the end as it can be done in one go for both cases. 
In the case $\gamma_+=\gamma_-$, relation \eqref{relation_a1} implies that if $\lda_k$ is a zero of $a(\lda)$ then so is $-\lda_k^*$ and the first claim follows. Note that none of these zeros can be equal to $(\beta+i\alpha)/2$ in view of the summary given after Lemma \ref{propBT}. Using \eqref{dressing_Psi} and \eqref{sym_psihat}, we obtain 
\begin{equation}
\label{Psi_B}
    \Psi_+(-x,-\lda)=B(x,\lda)\Psi_-(x,\lda)\cB^{-1}(\lda)
\end{equation} 
where $B(x,\lda)$ is given as in \eqref{folded_B} and $\cB(\lda)$ as in \eqref{BT2relationS}. This implies $\Psi_-(0,0)S(0)=B(0,0)\Psi_-(0,0)\cB^{-1}(0)$ \ie, since $B(0,0)=\1$ and $\cB(0)=\1$ when $\gamma_+=\gamma_-$, $S(0)=\1$. Thus $a(0)=1$. But from \eqref{form_a1}, we have $a(0)=(-1)^{2p+s}$ so $s$ must be even.

In the case $\gamma_+=-\gamma_-$, relation \eqref{relation_a2} is rephrased by introducing
\begin{equation}
    A(\lda)=\begin{cases}
    \displaystyle
    \frac{2\lda+\beta+i\alpha}{2\lda+\beta-i\alpha}\ a(\lda)\,,~~\varepsilon=1\,,\\
    \displaystyle
    \frac{2\lda-\beta+i\alpha}{2\lda-\beta-i\alpha}\ a(\lda)\,,~~\varepsilon=-1\,,
    \end{cases}
\end{equation}
in terms of which it reads $A(-\lda)=A^*(\lda^*)$. This is the same relation as dealt with before so we deduce that the zeros of $A(\lda)$ come in pairs or in singlets of purely imaginary numbers and \eqref{form_a2} follows. We use again \eqref{Psi_B} to deduce 
$\Psi_-(0,0)S(0)=B(0,0)\Psi_-(0,0)\cB^{-1}(0)$ but this time, since $\gamma_+=-\gamma_-=\varepsilon\alpha$,  $\cB^{-1}(0)=\begin{pmatrix}
    \frac{\beta+i\varepsilon\alpha}{\beta-i\varepsilon\alpha} & 0\\
    0 & \frac{\beta-i\varepsilon\alpha}{\beta+i\varepsilon\alpha}
\end{pmatrix}$. Therefore, $a(0)=\frac{\beta-i\varepsilon\alpha}{\beta+i\varepsilon\alpha}$. Comparing with \eqref{form_a2} which gives $a(0)=(-1)^{2p+s}\frac{\beta-i\varepsilon\alpha}{\beta+i\varepsilon\alpha}$, we deduce again that $s$ is even. In fact $s=0$ because \eqref{no_self} (whose proof is next) would implies $|\gamma(i\sigma_k)|^2=-1$, a contradiction.

We turn to the proof of the relation on the norming constants for which we don't need to distinguish the cases. Eq. \eqref{Psi_B} gives the following relations between the column vectors of $\Psi_\pm$
\small
\begin{eqnarray}
\label{Psi11}
&&    \Psi_+^{(1)}(-x,-\lda)=B(x,\lda)\Psi_-^{(1)}(x,\lda)\frac{-2\lda+\rho+i\gamma_+}{2\lda+\rho+i\gamma_-}\,,\\
\label{Psi22}
&&\Psi_+^{(2)}(-x,-\lda)=B(x,\lda)\Psi_-^{(2)}(x,\lda)\frac{-2\lda+\rho-i\gamma_+}{2\lda+\rho-i\gamma_-}\,.
\end{eqnarray}
\normalsize
Evaluating \eqref{Psi11} at $\lda=\lda_k$ and recalling that $\Psi_-^{(1)}(x,\lda_k)=\gamma(\lda_k)\Psi_+^{(2)}(x,\lda_k)$ we get
$$\Psi_+^{(1)}(-x,-\lda_k)=\gamma(\lda_k)B(x,\lda_k)\Psi_+^{(2)}(x,\lda_k)\frac{-2\lda_k+\rho+i\gamma_+}{2\lda_k+\rho+i\gamma_-}\,.$$
Now, evaluating \eqref{Psi22} at $\lda=-\lda_k$ and using it to eliminate $\Psi_+^{(2)}(x,\lda_k)$ yields
$$\Psi_+^{(1)}(-x,-\lda_k)=\gamma(\lda_k)\underbrace{B(x,\lda_k)B(-x,-\lda_k)}_{\1}\Psi_-^{(2)}(-x,-\lda_k)\frac{2\lda_k+\rho-i\gamma_+}{-2\lda_k+\rho-i\gamma_-}\frac{-2\lda_k+\rho+i\gamma_+}{2\lda_k+\rho+i\gamma_-}\,.$$
Recalling the NLS symmetry \eqref{waveFunction_NLS_symm} which gives
$$\Psi_+^{(1)}(x,\lda)=-N\Psi_+^{(2)*}(x,\lda^*)\,,~~\Psi_-^{(2)}(x,\lda)=N\Psi_-^{(1)*}(x,\lda^*)\,,$$
and the relation $\Psi_-^{(1)*}(-x,-\lda_k^*)=\gamma^*(-\lda_k^*)\Psi_+^{(2)*}(-x,-\lda_k^*)$, we get
$$\Psi_+^{(2)*}(-x,-\lda_k^*)=-\gamma(\lda_k)\gamma^*(-\lda_k^*)\frac{2\lda_k+\rho-i\gamma_+}{-2\lda_k+\rho-i\gamma_-}\frac{-2\lda_k+\rho+i\gamma_+}{2\lda_k+\rho+i\gamma_-}\Psi_+^{(2)*}(-x,-\lda_k^*)\,.$$
The result follows by spelling out the two cases $\gamma_+=\gamma_-$ or $\gamma_+=-\gamma_-$ and substituting $\rho=\beta$ and $\gamma_+=\varepsilon\alpha$.
\finprf

The main novelty is the first fraction in \eqref{form_a2} which shows the presence of a single, not purely imaginary, zero. In the absence of any other zero \ie $n=0$, and assuming the pure soliton case \ie $b(\lda)=0$, this is precisely the term that gives rise to the emitted or absorbed soliton in \eqref{eq:u1}. The norming constant related to this zero has no special constraint, which explains that $x_0$ and $\phi$ are arbitrary in \eqref{eq:u1}.

\begin{rem}
In expression \eqref{form_a2}, we could ask what happens for instance if one zero $\lda_k$ is equal to $(-\beta+i\alpha)/2$. 
A short calculation shows that $a(\lda)$ would then contain the factor $\left(\frac{2\lda+\beta-i\alpha}{2\lda+\beta+i\alpha}\right)^2\frac{2\lda-\beta-i\alpha}{2\lda-\beta+i\alpha}$. In that case, $(-\beta+i\alpha)/2$ would be a double zero which takes us beyond our working hypothesis of generic potentials. However, this is a tantalising possibility in general to investigate the properties of emission and absorption of more elaborate soliton solutions by a time-dependent boundary. 
\end{rem}

We now discuss the converse of Propositions \ref{Relat_Q_QtildeBT2} and \ref{prop_sym_discrete}. The short argument in \cite{bikbaev1991initial} uses the known one-to-one correspondence in ISM between a generic potential of the type we consider in this work and its scattering data. We can invoke the same result here and conclude that a potential $u$ is such that $\widehat{u}$ is odd (equivalently $\widetilde{u}(x)=-u(-x)$) if and only if its scattering data satisfies the symmetries of Propositions \ref{Relat_Q_QtildeBT2} and \ref{prop_sym_discrete}. In the case $\gamma_+=\gamma_-$, we also present in Appendix \ref{proof_main_result} a direct (but long) proof along the lines of that given in \cite{deift2011long} which uses Riemann-Hilbert problems techniques. It illustrates the main differences between the present case and the Robin case detailed in \cite{deift2011long}. In particular the use of a two-step construction mimicking the construction of $B(x,\lda)$ from $L_1(x,\lda)$ is detailed.
\begin{prop}\label{main_prop}
	Let $Q(x)\in\cS(\RR)$ be such that its scattering data satisfies the symmetries of Propositions \ref{Relat_Q_QtildeBT2} and \ref{prop_sym_discrete}. Then $\widetilde{u}(x)=-u(-x)$ holds.
\end{prop}
The compatibility of this whole construction with the desired time evolution must be established. Since $B(x,\lda)$ is constructed entirely on $L_1(x,\lda)$ (composed as in \eqref{folded_B}), this step is ensured by Lemma \ref{time_evol} and the setup explained before it. Specifically
\begin{prop}\label{Id_B_at_zero}
For each $t\ge 0$, let $u(x,t)\in \mathcal{S}(\mathbb{R})$ be a given solution of NLS, and $L_1(x,t,\lda)$ be as in \eqref{choiceL} with $P(x,t)$ constructed as in Lemma \ref{propBT}. Suppose $\widehat{u}(x,t)=L_{1\rho,\gamma_+}[u](x,t)$ is and odd function in $x$. Then, with $\rho=\beta$, $\gamma_+=\varepsilon\alpha$,
\be
B(0,t,\lda)=K(t,\lda)\,,
\ee
and 
\be
K_t(t,\lambda)= V(0,t,-\lambda)  K(t,\lambda)- K(t,\lambda) V(0,t,\lambda)\,.
\ee
\end{prop}
\prf Lemma \ref{time_evol} applied to $B(x,t,\lda)$, combined with the fact that $\widetilde{V}(x,t,\lda)=V(x,t,-\lda)$ (since $\widehat{u}(x,t)=-\widehat{u}(-x,t)$ is equivalent to $\widetilde{u}(x,t)=-u(-x,t)$), yields 
$$B_t(0,t,\lda)=V(0,t,-\lambda)  B(0,t,\lda)- B(0,t,\lda) V(0,t,\lambda)\,.$$
It remains to show that $B(0,t,\lda)=K(t,\lda)$. This is exactly the same calculation leading to \eqref{folded_B_at_origin}-\eqref{matrixH} but with the time dependence included. For completeness, here are the main steps.
A direct calculation yields 
\begin{eqnarray}
    \label{B_in_terms_P}
    &&B(x,t,\lda)\\
    &&=h(\lda)\left[-4\lda^2\1-4\lda\sigma_3\left(P(x,t)+P(-x,t)\right)+2\rho\sigma_3\left(P(x,t)-P(-x,t)\right)-4\sigma_3P(x,t)\sigma_3P(-x,t)+\rho^2\right]\nonumber\,,
\end{eqnarray}
so
\be
B(0,t,\lda)=h(\lda)\left[-4\lda^2-8\lda\sigma_3P(0,t)-4(\sigma_3P(0,t))^2+\rho^2\right]\,.
\ee
For each $t\ge 0$, $P(x,t)$ has the properties given in Proposition \ref{propP} so, recalling that $\widehat{u}(x,t)$ is an odd function in $x$ (so $\widehat{u}(0,t)=0$), we get 
\be
\sigma_3P(0,t)=\frac{i}{2}\begin{pmatrix}
	\epsilon(0,t) \sqrt{\gamma_+^2-|u|^2(0,t)}& u(0,t)\\
	 u^*(0,t) & -\epsilon(0,t) \sqrt{\gamma_+^2-|u|^2(0,t)}
\end{pmatrix}\,.
\ee
This also gives $(\sigma_3P(0,t))^2=-\frac{\gamma_+^2}{4}\1$, completing the proof. 
\finprf
The result that the symmetries on the scattering data are compatible with the time evolution also holds by the same reasoning that $B$ is the composition of two B\"acklund transformations constructed on $L_1$. 

So far, the method outlined gives a way to construct solutions of the desired initial-boundary value problem from solutions on the full line with special symmetries. This already provides a very large class of solutions to the initial-boundary value problem, including our new class of solutions with one absorbed/emitted soliton. Strictly speaking, as in the Robin case, the final step to claim to have solved the original initial-boundary value problem is to consider an appropriate extension of the initial data given on the half-line to an admissible initial data on the full line which automatically satisfies the desired folding symmetry. In \cite{deift2011long}, it was shown that, in the Robin case, all solution of the initial-boundary value problem are actually of this form, meaning that they arise as restrictions of solutions on the full line obtained by extension and having the appropriate symmetries. In the present paper, we do not aim at proving the analogous general result. Let us finish this section by presenting some results about this final step. The above construction ensures that we can follow the prescription of \cite{bikbaev1991initial}: if $u\in\cS(\RR^+)$ is the given data on the half-line and $\widetilde{u}(x)$ its B\"acklund transform under $B$, then set 
\begin{equation}
	\label{ext1}
	u^{ext}(x)=\begin{cases}
	u(x)\,,~~x> 0\,,\\
	-\widetilde{u}(-x)\,,~~x<0\,.
	\end{cases}
	\end{equation}
This extension satisfies the desired symmetry property by construction since $B$ induces an involution. The only technical point is its smoothness at $x=0$. As in the Robin case, the boundary conditions ensure continuity of the function, its first derivative and its second derivative automatically. In the present case we have the additional results that all higher odd order derivative are also continuous. The continuity of the even ones could be ensured in principle by imposing higher order boundary conditions in the spirit of Appendix D in \cite{bikbaev1991initial}. In fact, the recent work \cite{zhang2021inverse} gives an account on such higher boundary conditions which are compatible with NLS. We do not elaborate further on this here and simply give the following partial result on this issue.
\begin{lemma}[Smoothness of the B\"acklund extension]
The B\"acklund extension \eqref{ext1} is of class $C^2$ and its odd derivatives to all orders are continuous.
\end{lemma}
\prf
By definition, $Q^{ext}(x)$ is an element of $\mathcal{S}\left(\mathbb{R}\backslash\{0\}\right)$. We only need to check its smoothness properties at $x=0$. For convenience, let us write \eqref{B_in_terms_P} for short as $B(x,\lda)=h(\lda)\left[-4\lda^2+\lda B_1(x)+B_2(x)\right]$, where \small$B_1(x)=-4(\sigma_3P(x)+\sigma_3P(-x))$ and $B_2(x)=2\rho(\sigma_3P(x)-\sigma_3P(-x))-4\sigma_3P(x)\sigma_3P(-x)-\rho^2$. \normalsize From \eqref{EqB} it follows that 
\begin{equation}
\widetilde{Q}(x)=Q+\frac{i}{4}[B_1,\sigma_3],\quad  B_{1x}=i[B_2,\sigma_3]+\widetilde{Q}B_1-B_1Q,\quad  B_{2x}=\widetilde{Q}B_2-B_2Q.
\end{equation}
A direct calculation shows that $Q^{ext}(0^+)-Q^{ext}(0^-)=2Q(0)+2i[\sigma_3,\sigma_3P(0)]=0$, which means that $Q^{ext}(x)$ is continuous at $x=0$. Using the explicit expression of $B_1(x)$ above, we deduce that $\partial_{2n+1}B_1(0)=0$, for all positive integer $n$. Therefore, all $\partial_{2n+1}Q^{ext}(x)$ are exist and are continuous at $x=0$. Finally,
\small
\begin{align}
    Q_{xx}^{ext}(0^+)-Q_{xx}^{ext}(0^-)&=Q_{xx}(0)+\widetilde{Q}_{xx}(0)\nonumber\\
                                       &=2Q_{xx}(0)+\frac{i}{2}[B_{1xx}(0),\sigma_3]\nonumber\\
                                       &=2\begin{pmatrix}
                                           0& u_{xx}+(\alpha^2+\beta^2)u-2\varepsilon(0)u_x\Lambda\\
                 -u^*_{xx}-(\alpha^2+\beta^2)u^*+2\varepsilon(0)u_x^*\Lambda&0
                                       \end{pmatrix}\nonumber\\
                                       &=0,
\end{align} 
\normalsize
where $\Lambda=\sqrt{\alpha^2-|u|^2(0)}$, on account of the fact that we assume that $u$ satisfies \eqref{BCQ}.
\finprf
These smoothness properties are compatible with time evolution by construction.

\subsection{Special case: multisoliton solutions}\label{solitons}

A special case of our results contains those in \cite{gruner2020dressing} that were obtained by dressing. Specifically, the formulas for the position and phase shifts presented in Remark 2 of \cite{gruner2020dressing} can be obtained from the formulas in Proposition \ref{prop_sym_discrete}, in the case $\gamma_+=\gamma_-$, with some standard algebraic manipulations. Their explicit form in general is not crucial for our purposes. We will see an example below. In structure, these are the same as the ones originally given in \cite{biondini2009solitons} for the Robin case. The essential difference accounting for the presence of different boundary conditions is the appearance of the function  $\frac{2\lda+\beta-i\varepsilon\alpha}{2\lda+\beta+i\varepsilon\alpha}\frac{2\lda-\beta-i\varepsilon\alpha}{2\lda-\beta+i\varepsilon\alpha}$ which replaces the function $\frac{2\lda-i\theta}{2\lda+i\theta}$ characteristic of the Robin case. However, the new case $\gamma_+=-\gamma_-$ has not been seen before by the method of \cite{gruner2020dressing}.

\paragraph{Two solitons reflected. }

In the case $\gamma_+=\gamma_-$, we can apply our results to compute a two-soliton solution on the half-line being reflected by the boundary at $x=0$. It suffices to use the four-soliton solution of NLS on the full line, recalled in Section \ref{sec:ms}, with 
the constraints given in Proposition \ref{prop_sym_discrete}: the discrete data
satisfies $\lambda_3=-\lambda_1^*$, $\lambda_4=-\lambda_2^*$
and the associated norming constants are linked by, for k=1,2:
\begin{equation}\label{BT2relationnormingC}
   c(\lda_{k+2})^*=\frac{-1}{ c(\lda_k) a^\prime(\lda_k)a^\prime(-\lda_k^*)^*}\ 
    \frac{2\lda_k-\beta+i\varepsilon\alpha}{2\lda_k-\beta-i\varepsilon\alpha} \ \frac{2\lda_k+\beta+i\varepsilon\alpha}{2\lda_k+\beta-i\varepsilon\alpha} \ .
\end{equation}

\begin{figure}[htp]
    \centering
    \includegraphics[width=6cm]{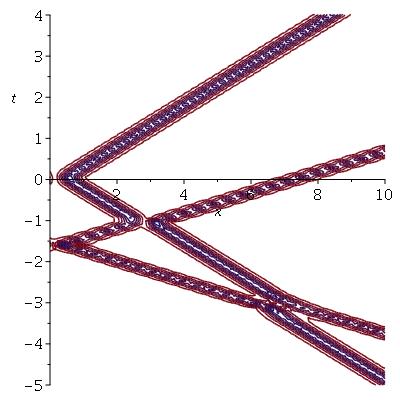}
    \includegraphics[width=6cm]{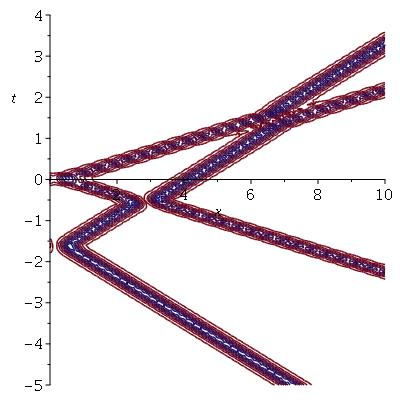}
    \caption{2D-contour plots of $|u(x,t)|$ corresponding to two solitons reflected with time-dependent BCs \eqref{BCQn} for $\alpha=2$ and $\beta=1$. The same zeros $\lambda_1=1+2i$ and $\lambda_2=(1+5i)/2$ are used for both plots and with $\varepsilon=1$ in \eqref{BT2relationnormingC}.
    The norming constants are $c(\lambda_1)= -4e^{-20},\  c(\lambda_2)=5e^{5}$ on the left and $c(\lambda_1)= -4e^{4},\  c(\lambda_2)=5e^{-10}$ on the right.}
    \label{plot_2soliton}
\end{figure}

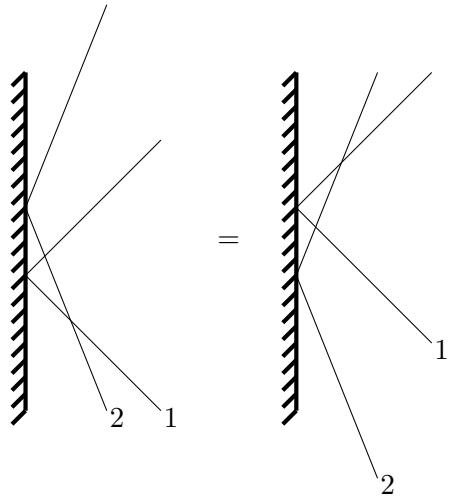
\begin{figure}[h!]
\begin{center}
  \begin{tikzpicture}[scale=0.9]
  \foreach \n in {0,-0.25,...,-5} {
  \draw[ultra thick] (0,\n)--(-0.2,\n-0.2);}
   \foreach \n in {0,-0.25,...,-5} {
  \draw[ultra thick] (4,\n)--(4-0.2,\n-0.2);}
  \draw[ultra thick] (0,0)--(0,-5);
  \draw (0,-3)--(2,-5); \draw (0,-3)--(2,-1);
  \draw (0,-2)--(1.2,-5); \draw (0,-2)--(1.2,1);
  \node at (3,-2.5) {$=$}; 
   \draw[ultra thick] (4,0)--(4,-5);
  \draw (4,-2)--(6,-4); \draw (4,-2)--(6,0);
  \draw (4,-3)--(5.2,-6); \draw (4,-3)--(5.2,0);
  \node at (2.15,-5.1) {$1$}; \node at (1.35,-5.1) {$2$};
  \node at (6.15,-4.1) {$1$}; \node at (5.35,-6.1) {$2$};
   \end{tikzpicture}
\end{center}
	\caption{Line representation of the (quantum) reflection equation depicting a two-particle process being factorised into two possible 
	successions of particle-particle interactions and particle-boundary interactions. The consistency of the two possibilities which must yield the same physical scattering matrix requires the reflection equation.}\label{ref_eq}
\end{figure}

 Fig.~\ref{plot_2soliton} shows two plots of two-soliton solutions for different choices of the parameters and both reflected by the boundary. 
Of course, such pictures will look very familiar to the reader accustomed to solitons reflections in the Robin case. The point is to offer a visual appreciation of the integrability of the time-dependent case. Between the two plots, the only parameter that we changed  
is the position shift $\xi_1$, $\xi_2$ of each soliton. This is the analog with a boundary of the well known property that solitons undergo elastic collisions whose order is irrelevant on the final result of position and phase shifts. The plots in Fig.~\ref{plot_2soliton} are to be compared with the graphical representation of the (quantum) reflection equation in Fig.~\ref{ref_eq} which is well known in quantum integrable systems. The main reason to mention this is that, in the multicomponent case, soliton collisions among themselves and with a boundary are related with the set-theoretical Yang-Baxter and reflection equations and provide examples of Yang-Baxter and reflection maps, see \cite{caudrelier2014zhang} and references therein. The extension of such ideas to the present time-dependent BCs is an interesting open problem. 

\paragraph{One soliton reflected and one absorbed.}
\begin{figure}[htp]
    \centering
    \includegraphics[width=6cm]{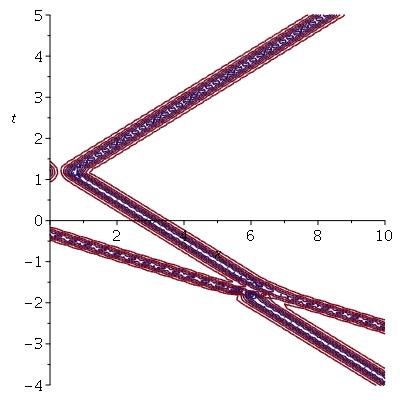}
    \includegraphics[width=6cm]{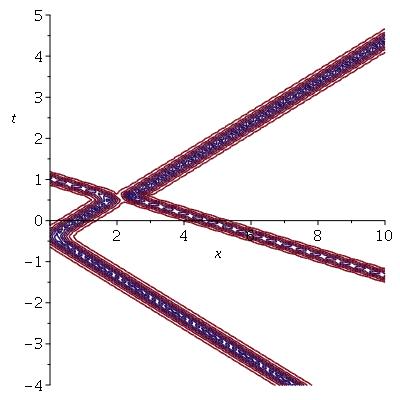}
    \caption{2D-contour plots of $|u(x,t)|$ corresponding to two solitons, one reflected and one absorbed with time-dependent BCs \eqref{BCQn} for $\alpha=4$, $\beta=2$. The same zeros $\lambda_0=1+2i$ and $\lambda_1=(1+5i)/2$ are used for both plots.
    The norming constants are
    $c(\lambda_0)= 1,\ c(\lambda_1)= 5e^{15}$ on the left and 
    $c(\lambda_0)= e^{20},\ c(\lambda_1)= 5$ on the right.}
    \label{plot_2soliton_absorbed}
\end{figure}

\begin{figure}[htp]
    \centering
    \includegraphics[width=6cm]{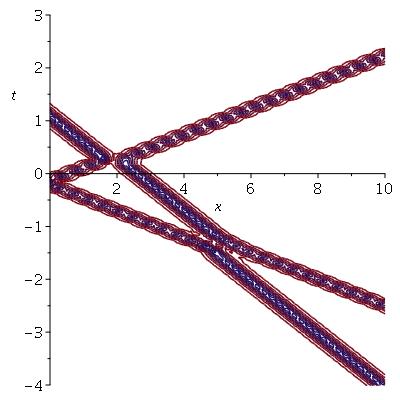}
    \includegraphics[width=6cm]{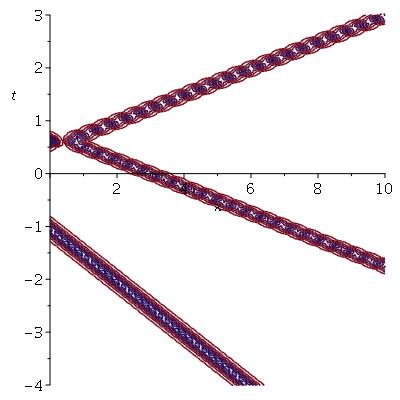}
    \caption{2D-contour plots of $|u(x,t)|$ corresponding to two solitons, one reflected and one absorbed with time-dependent BCs \eqref{BCQn} for $\alpha=6$, $\beta=1$. The same zeros $\lambda_0=(1+6i)/2$ and $\lambda_1=(2+5i)/2$ are used for both plots.
    The norming constants are
    $c(\lambda_0)= 4e^{16},\ c(\lambda_1)= 5$ on the left and 
    $c(\lambda_0)= 4e^{-8},\ c(\lambda_1)= 5e^{15}$  on the right.}
    \label{plot_2soliton_absorbedb}
\end{figure}

The previous result with two-soliton reflected on the half-line is similar to the ones obtained for the Robin boundary condition and corresponds to the case $\gamma_+=\gamma_-$. We now turn to the new possibility offered by the time-dependent case \ie $\gamma_+=-\gamma_-$. A completely new type of solution is then possible, as mentioned previously: one soliton may be absorbed/emitted by the boundary. To illustrate this type of solutions, in this paragraph, we focus on the case when one soliton is absorbed and one is reflected. This is obtained from a three-soliton solution of NLS on the full line, recalled in Section \ref{sec:ms}, with $a(\lda)$ given by \eqref{form_a2} with $p=1$ (and $b(\lambda)=0$). This means that two of the three zeros/norming constant are required to obey the symmetry relations of Proposition \ref{prop_sym_discrete} part 2. The third zero, say $\lda_0$ is the special zero involving the boundary parameters $\alpha$ and $\beta$: $\lda_0=-\beta/2+i\alpha/2$ or $\lda_0=\beta/2+i\alpha/2$. The sign of the real part determines whether the soliton travels towards or away from the boundary. 
The norming constants associated to the reflected soliton are linked by 
(see \eqref{no_self})
\begin{equation}\label{BT2relationnormingC2}
   c(\lda_{2})=\frac{-1}{ c(\lda_1)^*\,  a^\prime(\lda_1)^*\, a^\prime(-\lda_1^*)}\ ,
\end{equation}
whereas the norming constant $c(\lda_{0})$ is free.

Fig.~\ref{plot_2soliton_absorbed} and Fig.~\ref{plot_2soliton_absorbedb} show such plots for different choices of parameters, to mimick the situation in Fig.~\ref{plot_2soliton}, but with the essential difference that one of the two incoming soliton is absorbed by the boundary, while the other is reflected as before. 

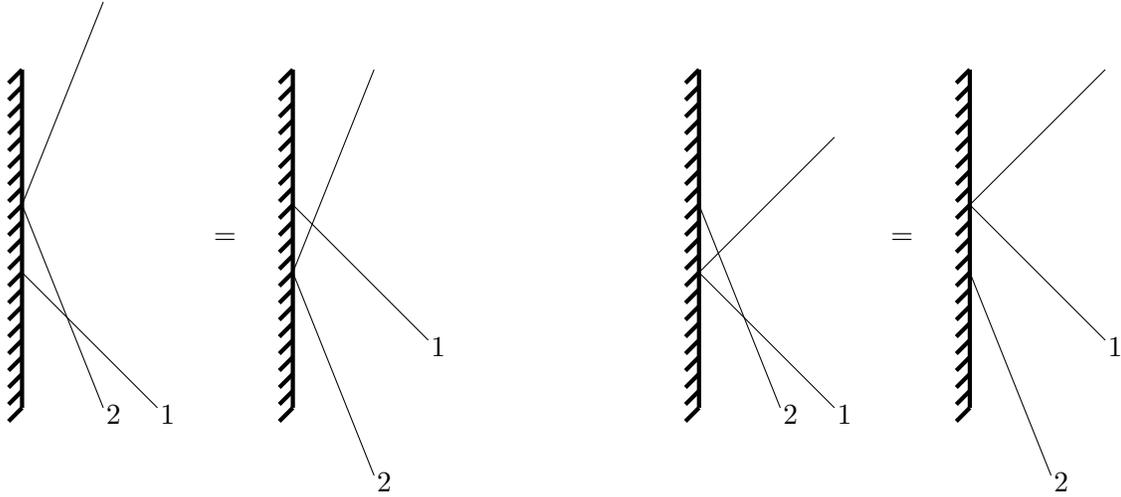
\begin{figure}[h!]
\begin{center}
  \begin{tikzpicture}[scale=0.9]
  
  \foreach \n in {0,-0.25,...,-5} {
  \draw[ultra thick] (0,\n)--(-0.2,\n-0.2);}
   \foreach \n in {0,-0.25,...,-5} {
  \draw[ultra thick] (4,\n)--(4-0.2,\n-0.2);}
  \draw[ultra thick] (0,0)--(0,-5);
  \draw (0,-3)--(2,-5); 

  \draw (0,-2)--(1.2,-5); \draw (0,-2)--(1.2,1);
  \node at (3,-2.5) {$=$}; 
   \draw[ultra thick] (4,0)--(4,-5);
  \draw (4,-2)--(6,-4); 

  \draw (4,-3)--(5.2,-6); \draw (4,-3)--(5.2,0);
  \node at (2.15,-5.1) {$1$}; \node at (1.35,-5.1) {$2$};
  \node at (6.15,-4.1) {$1$}; \node at (5.35,-6.1) {$2$};

    \foreach \n in {0,-0.25,...,-5} {
  \draw[ultra thick] (10,\n)--(10-0.2,\n-0.2);}
   \foreach \n in {0,-0.25,...,-5} {
  \draw[ultra thick] (14,\n)--(14-0.2,\n-0.2);}
  \draw[ultra thick] (10,0)--(10,-5);
  \draw (10,-3)--(12,-5); \draw (10,-3)--(12,-1);
  \draw (10,-2)--(11.2,-5); 
 
  \node at (13,-2.5) {$=$}; 
   \draw[ultra thick] (14,0)--(14,-5);
  \draw (14,-2)--(16,-4); \draw (14,-2)--(16,0);
  \draw (14,-3)--(15.2,-6); 
 
  \node at (12.15,-5.1) {$1$}; \node at (11.35,-5.1) {$2$};
  \node at (16.15,-4.1) {$1$}; \node at (15.35,-6.1) {$2$};
   \end{tikzpicture}
   
\end{center}
	\caption{Line representation of two-soliton solutions when one is absorbed.}\label{ref_eqn1}
\end{figure}

Fig. \ref{ref_eqn1} shows the line representation of the equations underpinning the phenomenon of Fig.~\ref{plot_2soliton_absorbed} and Fig.~\ref{plot_2soliton_absorbedb}, in the same way as Fig.~\ref{ref_eq} does for Fig.~\ref{plot_2soliton}. Somewhat intriguingly, if we interpret the absorption (or emission) of the single soliton as a type of transmission into the boundary (or to the mirror half-line), these equations correspond to (quantum) reflection-transmission equations, see e.g. \cite{caudrelier2005reflection,caudrelier2005factorization} and references therein. This puzzling observation deserves further investigation beyond the scope of this work.

\paragraph{One soliton reflected and one emitted. } 

A solutions with one soliton reflected and one emitted can simply be obtained from the previous solution by changing the sign of the velocity of the absorbed soliton in the previous paragraph \ie the sign of the real part of $\lda_0$. Graphs for such solutions are easily 
obtained by reversing the time flow, $t\rightarrow -t$, in the figures of the previous paragraph.

\section{Conclusions and outlook}\label{conclusions}
Motivated by a programme of unification of various approaches to integrable boundary conditions, we implemented the nonlinear mirror image method to construct solutions for NLS on the half-line with an time-dependent boundary condition at $x=0$. The main technical difficulty was to overcome the time-dependence of the B\"acklund matrix at $x=0$ by imposing its behaviour at infinity instead. We showed that in the Robin case, this provides a completely equivalent procedure as the well-known one orginally developed in \cite{bikbaev1991initial}. We then applied it to the new, time-dependent case to successfully map the initial-boundary value problem to a problem on the full line with certain symmetries on the scattering data. Doing so, in addition to a class of solutions that is very similar in structure to those in the Robin case, we discovered a class of solutions that could not be seen in the time-independent case. In terms of soliton content, this class predicts that there is a single soliton that can be absorbed or emitted by the boundary while the others are paired up as in the Robin case. A mechanism of soliton generation by a boundary was identified in \cite{fokas1992soliton} with methods of the unified transform. In this formalism, conditions for such solutions were formulated in the spectral space only and soliton generation was derived using the long-time asymptotic techniques of the nonlinear steepest descent method. In particular, it is not know what boundary conditions in real space would give rise to this behaviour. We stress that the present results are different in nature for two reasons: we are able to provide explicit formulas for the soliton absorption or emission and the boundary conditions in real space are clearly identified and integrable. 

As mentioned in the introduction, the programme of unification of approaches to boundary conditions in integrable PDEs has made some progress and, as the present work shows, leads to interesting new effects, but it is far from complete as far as the Fokas method is concerned. The missing link is the incorporation of {\it time-dependent linearizable BCs} in the Fokas method. From what we have explained, those are defined to be such that there exists a matrix $K(t,\lambda)$ and a transformation $\nu$ of the spectral parameter such that
\begin{equation*}
    \partial_t K(t,\lambda)=V(0,t,\lambda)\,K(t,\lambda)-K(t,\lambda)\,V(0,t,\nu(\lambda))\,.
\end{equation*}
In the case where $\nu$ is an involution (like in the present work where $\nu(k)=-k$, one can then asks the question of the connection with the nonlinear mirror image method, as was done in \cite{BiondiniFD2014} for Robin BCs. Beyond this, the case of (modified) KdV , which was one the main motivations behind the Fokas method, seems very interesting as $\nu$ is not an involution and a connection with a version of the mirror image method is not clear.

We should point out that we only investigated a single absorbed or emitted soliton in the new class of solutions since we had been working all along under the assumption of single zeros in the scattering coefficient $a(\lda)$ (generic potentials). However, by lifting this constraint, it is possible to imagine having solution corresponding to higher order zeros whose values are dictated by the boundary parameters, producing potentially trains of absorbed or emitted solitons or bound states of such solitons. This tantalising possibility deserves further investigation. 

Related to this last point, the question of ``higher'' compatible time-dependent boundary conditions, in the sense for instance of the recent work \cite{zhang2021inverse} is also an avenue for potentially interesting phenomena at the boundary. The presence of more boundary parameters could allow in principle to have solitons with different speed or amplitude being absorbed or emitted. This is speculation at this stage but appears to be an interesting open problem.

 As mentioned in Section \ref{solitons}, the generalisation of the notion of reflection maps introduced in \cite{caudrelier2014zhang} in relation with the vector NLS with time-independent boundary conditions seems a natural continuation of the present work with potentially two rewards: new examples of reflection maps and a completely new set of maps related to what could be called set-theoretical reflection-transmission maps. Again the latter point is rather speculative but we allow ourselves to mention it in the conclusion as this is rather intriguing possibility.
 
Finally, we have alluded to quantum counterparts of the present work in several places. Work is in preparation in order to understand some aspects of such a quantisation of time-dependent boundary conditions.

\paragraph{Acknowledgment:} N. Cramp\'e acknowledges the hospitality 
of the School of Mathematics, University of Leeds where this work was started. N. Cramp\'e's visit was partially supported by the Research Visitor Centre of the School of Mathematics.

\appendix

\section{Proof of Lemma \ref{propBT}}\label{proof_lemma}

It is convenient to work with $P_1(x)=P(x)\sigma_3$ which satisfies
\small
	\begin{equation}
    	\label{eqP1}
    	P_{1x}=\left[\frac{i\rho}{2}\sigma_3-Q+i\sigma_3 P_1,P_1\right]\,.
	\end{equation}
	\normalsize
This equation implies that $\frac{d}{dx}{\rm Tr}P_1^n=0$ for all $n\ge 1$. Hence, the eigenvalues of $P_1$ are constant and therefore equal to $\pm \frac{i\gamma_+}{2}$ in view of \eqref{Value_of_P} which translates into $\displaystyle \lim_{x\to+\infty}P_1(x)=\frac{i\gamma_+}{2}\sigma_3$. The eigenvalues being distinct, $P_1(x)$ is diagonalisable and we have 
\begin{equation*}
	P_1(x)=\varphi_1(x)\left(\frac{i\gamma_+}{2}\sigma_3\right)\varphi_1(x)^{-1}
\end{equation*}
for some invertible matrix $\varphi_1(x)$. Inserting back into \eqref{eqP1} yields
\small
\begin{equation*}
	\left[\varphi_{1x}(x)\varphi_1(x)^{-1},\varphi_1(x)\left(\frac{i\gamma_+}{2}\sigma_3\right)\varphi_1(x)^{-1}\right]=\left[\frac{i\rho}{2}\sigma_3-Q+i\sigma_3 \varphi_1(x)\left(\frac{i\gamma_+}{2}\sigma_3\right)\varphi_1(x)^{-1},\varphi_1(x)\left(\frac{i\gamma_+}{2}\sigma_3\right)\varphi_1(x)^{-1}\right]\,,
\end{equation*}
\normalsize
which means that 
\small
\begin{equation*}
	\varphi_{1x}(x)\varphi_1(x)^{-1}-\left(\frac{i\rho}{2}\sigma_3-Q+i\sigma_3 \varphi_1(x)\left(\frac{i\gamma_+}{2}\sigma_3\right)\varphi_1(x)^{-1}\right)=M(x)
\end{equation*}
\normalsize
with $\left[M(x),\varphi_1(x)\left(\frac{i\gamma_+}{2}\sigma_3\right)\varphi_1(x)^{-1}\right]=0$. In turn, this implies that $\varphi_1(x)^{-1}M(x)\varphi_1(x)\equiv D(x)$ is a diagonal matrix. The matrix $\varphi_1$ is not uniquely defined and it is always possible to consider the transformation $\varphi_1\mapsto \varphi_1h$ where $h$ is an invertible diagonal matrix without changing $P_1$. We use this freedom to choose $h$ such that $h_x=-Dh$ and set $\varphi=\varphi_1h$, with the conclusion that 
\small
\begin{equation*}
	P_1(x)=\varphi(x)\left(\frac{i\gamma_+}{2}\sigma_3\right)\varphi(x)^{-1}
\end{equation*}
\normalsize
where $\varphi$ is a nonsingular (or fundamental) solution of 
\begin{equation}
	\label{eq_varphi}
	\varphi_{x}(x)=\left(\frac{i\rho}{2}\sigma_3-Q\right)\varphi+i\sigma_3 \varphi(x)\left(\frac{i\gamma_+}{2}\sigma_3\right)\,.
\end{equation}
Writing $\varphi=(\varphi_1,\varphi_2)$ where $\varphi_{1,2}$ are the column vectors of $\varphi$, we see that 
\begin{equation*}
	\varphi_{1x}(x)=(-i\lambda_+\sigma_3-Q)\varphi_1\,,~~\varphi_{2x}(x)=(-i\lambda^*_+\sigma_3-Q)\varphi_2\,,~~\lda_+=-\frac{\rho+i\gamma_+}{2}\,.
\end{equation*}
We impose the standard conditions
\begin{equation}
	\label{lim_varphi}
	\lim_{x\to+\infty}\varphi_1(x)e^{i\lda_+x}=
	\begin{pmatrix}
		1\\
		0
	\end{pmatrix}\,,~~\lim_{x\to+\infty}\varphi_2(x)e^{-i\lda^*_+x}=
	\begin{pmatrix}
		0\\
		1
	\end{pmatrix}\,.
\end{equation}
Eq. \eqref{eq_varphi} is unchanged under the transformation $\varphi(x)\mapsto N\varphi^*(x)N^{-1}$ where 
$N=\begin{pmatrix}
	0 & -1\\
	1 & 0
\end{pmatrix}$. Hence, in general there exists a nonsingular matrix $C$ such that $\varphi(x)= N\varphi^*(x)N^{-1}C$. Taking into account \eqref{lim_varphi}, we find $C=\1$, so $\varphi_2(x)=N\varphi_1^*(x)$, and hence the matrix $\varphi(x)$ can be written as
$$\varphi(x)=
\begin{pmatrix}
	\xi_1(x) & -\xi_2(x)^* \\
	\xi_2(x) & \xi_1(x)^*
\end{pmatrix}\,.$$
Thus, $P_1(x)$ takes the following form
\small
\begin{equation}\label{deift_lemma_P_form}
	P_1(x)=\frac{i\gamma_+}{2(|\xi_1(x)|^2+|\xi_2(x)|^2)}
	\begin{pmatrix}
		|\xi_1(x)|^2-|\xi_2(x)|^2 & 2\xi_2(x)^*\xi_1(x)\\
		2\xi_1(x)^*\xi_2(x) & -\left(|\xi_1(x)|^2-|\xi_2(x)|^2\right).
	\end{pmatrix}
\end{equation}
\normalsize
Therefore, it is enough to solve for $\varphi_1$ which we recall is a solution of the linear problem
\bea
\label{diff_equat_phi1}
\varphi_{1x}=(-i\lda_+\sigma_3-Q)\varphi_1\,,~~
\eea
with the condition $\displaystyle \lim_{x\to+\infty}\varphi_1(x)e^{i\lda_+x}=e_1$. The rest of the construction of $P_1(x)$ and its properties hinges on the following important standard result, see e.g. \cite[pp. 104-105]{coddington1955theory}. 
Eq. \eqref{diff_equat_phi1} admits two fundamental solutions $\chi^\pm(x)$ satisfying
\begin{equation}
	\lim_{x\to\pm \infty}\chi^\pm(x)e^{i\lda_+x\sigma_3}=\1\,.
\end{equation}
Note that they are not necessarily unique. Also, we point out that their relation with the Jost solutions appropriately continued in the complex half-plane is key in the following. So we need to distinguish the cases $\gamma_+>0$ and $\gamma_+<0$. We will need the following known properties of the Jost solutions and scattering data (see e.g. \cite{gu2004darboux,petermiller2} for more details). For $\lda\in\mathbb{C}^\pm$,
\small
\begin{equation}\label{limR}
\lim_{x\to\mp\infty}\Psi_\mp^{(1)}(x,\lda)e^{i\lda x}=\begin{pmatrix}
    1\\0
\end{pmatrix}\,, \quad 
\lim_{x\to\pm\infty}\Psi_\pm^{(2)}(x,\lda)e^{-i\lda x}=\begin{pmatrix}
    0\\1
\end{pmatrix}\,,
\end{equation}
\normalsize
for $\lda$ in any compact subset of $\mathbb{C}^+$,
\small
\begin{equation}\label{limC1}
    \lim_{x\to+\infty}\Psi_-^{(1)}(x,\lda)e^{i\lda x}=\begin{pmatrix}
    a(\lda)\\
    0
\end{pmatrix}, \quad \lim_{x\to-\infty}\Psi_+^{(2)}(x,\lda)e^{-i\lda x}=\begin{pmatrix}
    0\\a(\lda)
\end{pmatrix},
\end{equation}
\normalsize
and for $\lda$ in any compact subset of $\mathbb{C}^-$,
\small
\begin{equation}\label{limC2}
    \lim_{x\to-\infty}\Psi_+^{(1)}(x,\lda)e^{i\lda x}= \begin{pmatrix}
    a^*(\lda)\\0
\end{pmatrix}\,,\quad \lim_{x\to+\infty}\Psi_-^{(2)}(x,\lda)e^{-i\lda x}=\begin{pmatrix}
    0\\a^*(\lda)
\end{pmatrix}.
\end{equation}
\normalsize
\paragraph{Case $\gamma_+<0$:} In that case $\lda_+\in\CC^+$ and it could be a zero of $a(\lda)$ or not. Suppose first that	
$a(\lda_+)\neq 0$. Then we know that $\Psi^{(1)}_-(x,\lda_+)$ and $\Psi^{(2)}_+(x,\lda_+)$ are linearly independent. 	
From the first equation in (\ref{a_Jost_sols}) and the asymptotic behaviour in \eqref{limR} and \eqref{limC1}, we see that we can take $X(x)=\sigma_3(\Psi^{(1)}_-(x,\lda_+)/a(\lda_+), \Psi^{(2)}_+(x,\lda_+))$ as a fundamental matrix for \eqref{diff_equat_phi1}. Hence, we have 
\begin{equation}\label{constant_mu}
	\varphi_1(x)=X(x)	\begin{pmatrix}
		\mu_1\\
		\mu_2
	\end{pmatrix}   
\end{equation}
for some constants $\mu_1$ and $\mu_2$. Condition \eqref{lim_varphi} yields $\mu_1=1$ but $\mu_2$ is free. 
As $x\to-\infty$, we have $\varphi_1(x)\sim\begin{pmatrix}
	e^{-i\lda_+x}/a(\lda_+)\\
	-\mu_2a(\lda_+)e^{i\lda_+x}
\end{pmatrix}$. There are two sub-cases, either $\mu_2\neq 0$ or $\mu_2=0$. Assume that $\mu_2\neq 0$. Then, $\frac{\xi_1(x)}{\xi_2(x)}\to 0$ as $ x\to-\infty$ 
which leads to 
$$\lim_{x\to-\infty}P_1(x)= \frac{-i\gamma_+}{2}\sigma_3\,.$$ 
If $\mu_2=0$, then
$$\lim_{x\to-\infty}P_1(x)= \frac{i\gamma_+}{2}\sigma_3\,.$$ 
Suppose now that $a(\lda_+)=0$. Then $\Psi^{(1)}_-(x,\lda_+)$ and $\Psi^{(2)}_+(x,\lda_+)$ are no longer linearly independent. Our strategy is to use the fundamental solution $\chi^-(x)$ to exhibit a convenient fundamental matrix that we will use to determine $\varphi_1(x)$. On the one hand, we know that $\varphi_1(x)=\chi^+(x)	\begin{pmatrix}
	1\\
	c
\end{pmatrix}$ for some constant $c$ and that $\chi^+(x)=\chi^-(x)C$ for some constant invertible matrix $C$. Hence $\varphi_1(x)=\chi^-(x)	\begin{pmatrix}
	\alpha\\
	\beta
\end{pmatrix}$ for some constants $\alpha,\beta$. Finally, we also have that $\sigma_3\Psi^{(1)}_-(x,\lda_+)=\chi^-(x)
\begin{pmatrix}
	1\\
	d
\end{pmatrix}$ for some constant $d$. Hence, let us define $Y(x)=(\sigma_3\Psi^{(1)}_-(x,\lda_+),\chi_2^-(x))$ where $\chi_2^-(x)$ is the second column vector of $\chi^-(x)$. This is also a fundamental matrix since $Y(x)=\chi^-(x)\begin{pmatrix}
	1 & 0\\
	d & 1
\end{pmatrix}$ and it satisfies 
\begin{equation*}
	\lim_{x\to- \infty}Y(x)e^{i\lda_+x\sigma_3}=\1\,.
\end{equation*}
Putting everything together, we have $\varphi_1(x)=Y(x)	\begin{pmatrix}
	\alpha\\
	\delta
\end{pmatrix}$ where $\delta=\beta-\alpha d$ is some constant. We now show that $\delta\neq 0$ necessarily. Since $a(\lda_+)=0$, we know that $\Psi^{(1)}_-(x,\lda_+)=\gamma\Psi^{(2)}_+(x,\lda_+)$ so that $Y(x)=(\gamma\sigma_3\Psi^{(2)}_+(x,\lda_+),\chi_2^-(x))$. Finally, we also have $\chi_2^-(x)=\chi^+(x)\begin{pmatrix}
	\mu\\
	\nu
\end{pmatrix}$ for some constants $\mu,\nu$. Hence,
\small
\begin{eqnarray}
	\lim_{x\to+\infty}\varphi_1(x)e^{i\lda_+x}
	&=&\lim_{x\to+\infty}\left(\gamma\sigma_3\Psi^{(2)}_+(x,\lda_+),\chi^+(x)\begin{pmatrix}
		\mu\\
		\nu
	\end{pmatrix} \right)\begin{pmatrix}
		\alpha\\
		\delta
	\end{pmatrix}e^{i\lda_+x}\nonumber\\
	&=&\lim_{x\to+\infty}\left(\gamma\sigma_3\Psi^{(2)}_+(x,\lda_+)e^{-i\lda_+x},\chi^+(x)e^{i\lda_+x\sigma_3}\begin{pmatrix}
		\mu\\
		\nu e^{2i\lda_+x}
	\end{pmatrix} \right)\begin{pmatrix}
		\alpha e^{2i\lda_+x}\\
		\delta
	\end{pmatrix}\nonumber\\
	&=&\begin{pmatrix}
		0 & \mu\\
		-\gamma & 0
	\end{pmatrix}\begin{pmatrix}
		0\\
		\delta
	\end{pmatrix}\,.
\end{eqnarray}
\normalsize
Comparing with \eqref{lim_varphi}, we obtain $\mu\delta=1$ thus showing that $\delta\neq 0$. Therefore, going back to $\varphi_1(x)=Y(x)	\begin{pmatrix}
	\alpha\\
	\delta
\end{pmatrix}$ with $Y(x)=(\sigma_3\Psi^{(1)}_-(x,\lda_+),\chi_2^-(x))$, we obtain that $\varphi_1(x)\sim\begin{pmatrix}
	\alpha e^{-i\lda_+x}\\
	\delta e^{i\lda_+x}
\end{pmatrix}$ as $x\to-\infty$, with $\delta\neq 0$. Hence, $\frac{\xi_1(x)}{\xi_2(x)}\to 0$ as $ x\to-\infty$ 
which leads to 
$$\lim_{x\to-\infty}P_1(x)= -\frac{i\gamma_+}{2}\sigma_3\,.$$ 
\paragraph{Case $\gamma_+>0$:} In that case $\lda_+\in\CC^-$ and it could be a zero of $a^*(\lda)$ or not, or equivalently, $\lda_+^*$ could be a zero of $a(\lda)$ or not. We follow a similar strategy as for the previous case but the change of sign in $\gamma_+$ yields a major difference: here $\varphi_1(x)=\sigma_3\Psi^{(1)}_+(x,\lda_+)$. Indeed, in general we have 
\begin{equation*}
	\varphi_1(x)=\chi^+(x)\begin{pmatrix}
		\nu_1\\
		\nu_2
	\end{pmatrix}   \,,~~\Psi^{(1)}_+(x,\lda_+)=\sigma_3\chi^+(x)\begin{pmatrix}
		\tau_1\\
		\tau_2
	\end{pmatrix}   \,,
\end{equation*}
for some constants $\nu_1$, $\nu_2$, $\tau_1$ and $\tau_2$. Imposing \eqref{lim_varphi} and the asymptotic of $\Psi_+^{(1)}(x,\lda)$ as in \eqref{limR} requires $\nu_1=1=\tau_1$ and $\nu_2=0=\tau_2$.
Suppose first that $a^*(\lda_+)\neq 0$, then we can use the first limit in \eqref{limC2} and deduce that 
$$\lim_{x\to-\infty}P_1(x)= \frac{i\gamma_+}{2}\sigma_3\,.$$ 
Suppose now that $a^*(\lda_+)= 0$ so that $\Psi^{(1)}_+(x,\lda_+)$ and $\Psi^{(2)}_-(x,\lda_+)$ are no longer linearly independent and $\Psi^{(1)}_+(x,\lda_+)=\gamma^{'}\Psi^{(2)}_-(x,\lda_+)$ for some nonzero constant $\gamma^{'}$. We can use the asymptotic behaviour of $\Psi_-^{(2)}(x,\lda)$ given in \eqref{limR} to conclude that 
$$\lim_{x\to-\infty}P_1(x)= -\frac{i\gamma_+}{2}\sigma_3\,.$$ 

This concludes the proof of part (a). For part (b), we can adapt the proof of \cite[Theorem 6.6]{TU1998poisson} whose main points are as follows. From the construction of $\varphi_1(x)$, in all cases, we have either $\xi_1(x)/\xi_2(x)$ or $\xi_2(x)/\xi_1(x)$ tends to $0$ exponentially as $e^{\mp |\gamma_+|x}$ as $x\to \pm\infty$. Hence, from \eqref{deift_lemma_P_form}, we see that $[\sigma_3,P_1(x)]$ decays exponentially as $x\to\pm\infty$. We can see that in the equation
\begin{equation}
	P_{1x}=\frac{i\rho}{2}[\sigma_3,P_1]+i[\sigma_3,P_1]P_1-[Q,P_1],
\end{equation}
the first two terms on the right-hand side decay exponentially while the third term has the same decay as $Q$ which is assumed to be in $\mathcal{S}(\mathbb{R})$. Thus $P_{1x}$ has the same decay as $Q$. Hence, by repeated differentiation of \eqref{eqP1}, we obtain that $P_{1x}$ has the same decay properties as $Q$ at $\pm\infty$ and therefore belongs to $\mathcal{S}(\mathbb{R})$.

\section{Proof of Proposition \ref{main_prop}}
\label{proof_main_result}
We consider the case $\gamma_-=\gamma_+$ and denote this common value by $\gamma$ instead of $\varepsilon\alpha$ for convenience in this proof. Similarly we keep $\rho$ instead of $\beta$. Let us assume that symmetries \eqref{relation_a1} and \eqref{norming-const1} are satisfied. We will adapt the strategy proposed in \cite{deift2011long} to show that $\widetilde{u}(x)=-u(-x)$ holds. Consider $\mu(x,\lda)=\left(\mu_1(x,\lda),\mu_2(x,\lda)\right)$ the solution to the normalized RHP with jump matrix given by 
\begin{equation*}
	v(x,\lda):=
	\begin{pmatrix}
		1+|r(\lda)|^2 & -r^*(\lda)e^{-2i\lda x}\\
		-r(\lda)e^{2i\lda x} & 1
	\end{pmatrix}.
\end{equation*}
Consider two functions $q_1(x)$ and $E_1(x)$ defined as
\begin{equation}
    q_1(x)=\sigma_3\left(\mu_1\left(x,\widehat{\lda}\right),\mu_2\left(x,\widehat{\lda}^*\right)\right),\, E_1(x)=q_1(x)\sigma_3\left(\frac{i\gamma}{2}\right)q_1(x)^{-1}\sigma_3,\quad \widehat{\lda}=-\frac{\rho+i\gamma}{2}.
\end{equation}
Set
\begin{equation}
    \widehat{\mu}(x,\lda)\equiv \begin{cases}
		Z(x,\lda)\mu(x,\lda)z(\lda)^{-1},\quad \lda\in\mathbb{C}^+\backslash\left\{\frac{-\rho+i|\gamma|}{2}\right\},\\
		Z(x,\lda)\mu(x,\lda)z(\lda)^{-1},\quad \lda\in\mathbb{C}^-\backslash\left\{\frac{-\rho-i|\gamma|}{2}\right\},
	\end{cases}
\end{equation}
where $Z(x,\lda)=\left(\lda+\frac{\rho}{2}\right)\sigma_3+E_1(x)$ and $z(\lda)=\left(\lda+\frac{\rho}{2}\right)\sigma_3+\frac{i\gamma}{2}\1$.
Consider other two functions $q_2(x)$ and $E_2(x)$
\begin{equation*}
	 q_2(x)=\left(\widehat{\mu}_1\left(x,-\widehat{\lda}\right),\widehat{\mu}_2\left(x,-\widehat{\lda}^*\right)\right),\quad E_2(x)=\sigma_3q_2(x)\left(\frac{i\gamma}{2}\right)\sigma_3 q_2(x)^{-1}.
\end{equation*}
Let us define the following matrix function
\begin{equation}\label{new_RHP}
	\widetilde{\mu}(x,\lda)\equiv \begin{cases}
		N W(x,\lda)	\widehat{\mu}(x,\lda)w(\lda)^{-1} a(\lda)^{\sigma_3}N^{-1},\quad \lda\in\mathbb{C}^+\backslash\left\{\frac{\rho+i|\gamma|}{2}\right\},\\
		N W(x,\lda)	\widehat{\mu}(x,\lda)w(\lda)^{-1}a^*(\lda^*)^{-\sigma_3}N^{-1},\quad \lda\in\mathbb{C}^-\backslash\left\{\frac{\rho-i|\gamma|}{2}\right\},
	\end{cases}
\end{equation}
where $W(x,\lda)=\left(-\lda+\frac{\rho}{2}\right)\sigma_3-E_2(x)$ and $w(\lda)=\left(-\lda+\frac{\rho}{2}\right)\sigma_3-\frac{i\gamma}{2}\1$. 

We claim:
\begin{equation}\label{claim}
	\mu(x,\lda)=\widetilde{\mu}^*(-x,-\lda^*).
\end{equation}
\footnotesize
\begin{align*}
	\left(\widetilde{\mu}_-\right)^{-1}\widetilde{\mu}_+(x,\lda) &= N a^*(\lda)^{\sigma_3}z(\lda)w(\lda)v(x,\lda)\left(z(\lda)w(\lda)\right)^{-1}a(\lda)^{\sigma_3}N^{-1}\\
	&= N\begin{pmatrix}
		(1+|r(\lda)|^2)|a|^2 & -\frac{2\lda+\rho+i\gamma}{2\lda-\rho+i\gamma}\frac{2\lda+\rho-i\gamma}{2\lda-\rho-i\gamma}\frac{a^*(\lda)}{a(\lda)}r^*(\lda)e^{-2i\lda x}\\
		-\frac{2\lda-\rho+i\gamma}{2\lda+\rho+i\gamma}\frac{2\lda-\rho-i\gamma}{2\lda+\rho-i\gamma}\frac{a(\lda)}{a^*(\lda)}r(\lda)e^{2i\lda x} & \frac{1}{|a(\lda)|^2}
	\end{pmatrix}N^{-1}\\
	&= \begin{pmatrix}
		1+|r(-\lda)|^2 &
		-e^{2i\lda x}r(-\lda)\\
		-e^{-2i\lda x}r^*(-\lda) & 1
	\end{pmatrix}.
\end{align*}
\normalsize
Set $\overline{\mu}(x,\lda)\equiv \widetilde{\mu}^*(-x,-\lda^*)$. Then, 
\begin{equation*}
	\left(\overline{\mu}_-\right)^{-1}\overline{\mu}_+(x,\lda)=\left(\left(\widetilde{\mu}_-\right)^{-1}\widetilde{\mu}_+(-x,-\lda)\right)^*=v(x,\lda).
\end{equation*}
When $a(\lda)$ admits a finite number of simple zeros $\lda_k\in \mathbb{C}^+$,
\begin{equation*}
	\underset{\lda=\lda_k}{\text{Res}}\mu(x,\lda)=\lim_{\lda\to \lda_k}(\lda-\lda_k)\mu(x,\lda)=\lim_{\lda\to \lda_k}\mu(x,\lda)\begin{pmatrix}
		0&0\\ c(\lda_k)e^{2i\lda_kx}&0
	\end{pmatrix}.
\end{equation*}
Equivalently,
\begin{equation*}
	\lim_{\lda\to \lda_k} a(\lda)\mu_1(x,\lda)=\gamma(\lda_k)e^{2i\lda_k x}\mu_2(x,\lda_k).
\end{equation*}
For $\overline{\mu}(x,\lda)$, we have
\small
\begin{align}\label{residue_mu_bar1}
	\left(\underset{\lda=-\lda_k^*}{\text{Res}} \overline{\mu}(-x,\lda)\right)^*& = \lim_{\lda\to -\lda_k^*}\left[(\lda-(-\lda_k^*))\overline{\mu}(-x,\lda)\right]^* =\lim_{\lda\to \lda_k}-(\lda-\lda_k)\overline{\mu}^*(-x,-\lda^*)\nonumber\\
	& =-\lim_{\lda\to \lda_k}N (WZ)(x,\lda)(\lda-\lda_k)\mu(x,\lda)a(\lda)^{\sigma_3}\left(z(\lda)w(\lda)\right)^{-1} N^{-1}\nonumber\\
	& = -N (WZ)(x,\lda_k)\left[0\quad \frac{\mu_2(x,\lda_k)}{a^\prime(\lda_k)}\right]\left((zw)(\lda_k)\right)^{-1}N^{-1}\nonumber\\
	& = -4N (WZ)(x,\lda_k)\left[\frac{1}{2\lda_k+\rho-i\gamma}\frac{1}{2\lda_k-\rho-i\gamma}\frac{1}{a^\prime(\lda_k)}\mu_2(x,\lda_k)\quad 0\right],
\end{align}
\normalsize
Thus at $-\lda_k^*$, the second column of $\overline{\mu}(x,\lda)$ is analytic, and the first column has a simple pole. Similarly, at $-\lda_k$ the first column of $\overline{\mu}(x,\lda)$ is analytic, and the second column has a simple pole. On the other hand, we have
\small
\begin{align}\label{residue_mu_bar2}
	\lim_{\lda\to -\lda_k^*}\left[\overline{\mu}(-x,\lda)\begin{pmatrix}
		0&0 \\ e^{-2i\lda_k^*x}\overline{c}(-\lda_k^*)&0
	\end{pmatrix}\right]^* & = \lim_{\lda\to \lda_k}\overline{\mu}^*(-x,-\lda^*)\begin{pmatrix}
		0&0 \\ e^{2i\lda_kx}\overline{c}^*(-\lda_k^*)&0
	\end{pmatrix}\nonumber\\
	& =\lim_{\lda\to \lda_k}N (WZ)(x,\lda)\mu(x,\lda)a(\lda)^{\sigma_3}\left[(zw)(\lda)\right]^{-1}\begin{pmatrix}
		e^{2i\lda_kx}\overline{c}^*(-\lda_k^*)&0 \\ 0&0
	\end{pmatrix}\nonumber\\
	&  =-4N (WZ)(x,\lda_k)\left[\frac{\gamma(\lda_k)}{2\lda_k+\rho+i\gamma}\frac{\overline{c}^*(-\lda_k^*)}{2\lda_k-\rho+i\gamma}\mu_2(x,\lda_k)\quad 0\right],
\end{align}
\normalsize
Note that $\overline{c}(\lda_k)$ stands for the discrete data that appears in the normalised RHP for $\overline{\mu}(x,\lda)$. Compare \eqref{residue_mu_bar1} and \eqref{residue_mu_bar2}, using \eqref{norming-const1} and the first equation in \eqref{relation_a1}, one gets
\small
\begin{align*}
	\overline{c}(-\lda_k^*)&=\left[\frac{2\lda_k-\rho+i\gamma}{2\lda_k-\rho-i\gamma}\frac{2\lda_k+\rho+i\gamma}{2\lda_k+\rho-i\gamma}\frac{1}{\gamma(\lda_k)}\frac{1}{a^\prime(\lda_k)}\right]^*=-\frac{\gamma(-\lda_k^*)}{(a^\prime(\lda_k))^*}=\frac{\gamma(-\lda_k^*)}{a^\prime(-\lda_k^*)}=c(-\lda_k^*).
\end{align*}
\normalsize
Hence, it follows that $\overline{c}(\lda_k)=c(\lda_k)$.
Let us discuss what happens at $\lda=\widehat{\lda},\widehat{\lda}^*,-\widehat{\lda},-\widehat{\lda}^*$:
\small
\begin{align*}
    \widehat{\mu}(x,\lda) & = Z(x,\lda)\mu(x,\lda)z(\lda)^{-1}
    =q_1(x)\sigma_3
    \begin{pmatrix}
	    \left[(\sigma_3q_1(x))^{-1}\mu\right]_{11} & -\frac{2\lda+\rho+i\gamma}{2\lda+\rho-i\gamma}\left[(\sigma_3q_1(x))^{-1}\mu\right]_{12}\\
	    -\frac{2\lda+\rho-i\gamma}{2\lda+\rho+i\gamma}\left[(\sigma_3q_1(x))^{-1}\mu\right]_{21} & \left[(\sigma_3q_1(x))^{-1}\mu\right]_{22}
    \end{pmatrix},
\end{align*}
\normalsize
Note that $\left[(\sigma_3q_1(x))^{-1}\mu(x,\lda)\right]_{21}=0$ and $\left[(\sigma_3q_1(x))^{-1}\mu(x,\lda)\right]_{12}=0$ at $\widehat{\lda}$ and $\widehat{\lda}^*$, respectively. This means that $\widehat{\mu}(x,\lda)$ does not have any pole at $\widehat{\lda}$ and $\widehat{\lda}^*$. Again
\small
\begin{align*}
    W(x,\lda)\widehat{\mu}(x,\lda)w(\lda)^{-1}&
    =\sigma_3q_2(x)\begin{pmatrix}
	\left[\sigma_3q_2(x)^{-1}\widehat{\mu}\right]_{11} & -\frac{2\lda-\rho+i\gamma}{2\lda-\rho-i\gamma}\left[\sigma_3q_2(x)^{-1}\widehat{\mu}\right]_{12}\\
	-\frac{2\lda-\rho-i\gamma}{2\lda-\rho+i\gamma}\left[\sigma_3q_2(x)^{-1}\widehat{\mu}\right]_{21} & \left[\sigma_3q_2(x)^{-1}\widehat{\mu}\right]_{22}
    \end{pmatrix}
\end{align*}
\normalsize
Note that $\left[\sigma_3q_2(x)^{-1}\widehat{\mu}\right]_{21}(x,-\widehat{\lda})=0$ and $\left[\sigma_3q_2(x)^{-1}\widehat{\mu}\right]_{12}(x,-\widehat{\lda}^*)=0$. Combined with the above discussion, we see that $\widetilde{\mu}^*(-x,-\lda^*)$ does not have extra poles at $\widehat{\lda},\widehat{\lda}^*,-\widehat{\lda},-\widehat{\lda}^*$. Therefore, we have proved the above claim.
In terms of vector columns, Eq \eqref{claim} is:
\newline
For $\lda\in \mathbb{C}^+$
\footnotesize
\begin{equation}\label{sym1}
	\mu_1(x,\lda)=\frac{-1}{a^*(-\lda^*)\left(-\left(\lda+\frac{\rho}{2}\right)+\frac{i\gamma}{2}\right)}N(W\widehat{\mu}_2)^*(-x,-\lda^*),\quad \mu_2(x,\lda)=\frac{a^*(-\lda^*)}{\lda+\frac{\rho}{2}+\frac{i\gamma}{2}}N(W\widehat{\mu}_1)^*(-x,-\lda^*),
\end{equation}
\normalsize
and, for $\lda\in \mathbb{C}^-$
\footnotesize
\begin{equation}\label{sym3}
	\mu_1(x,\lda)=\frac{-a(-\lda)}{-\left(\lda+\frac{\rho}{2}\right)+\frac{i\gamma}{2}}N(W\widehat{\mu}_2)^*(-x,-\lda^*),\quad \mu_2(x,\lda)=\frac{1}{a(-\lda)\left(\lda+\frac{\rho}{2}+\frac{i\gamma}{2}\right)}N(W\widehat{\mu}_1)^*(-x,-\lda^*)
\end{equation}
\normalsize
Assume that $\gamma<0$, so $\widehat{\lda}\in \mathbb{C}^+$. Using $N\widehat{\mu}^*(x,\lda^*)N^{-1}=\widehat{\mu}(x,\lda)$ and $NW^*(x,\lda^*)N^{-1}=W(x,\lda)$, it follows from the first Eq. in \eqref{sym1} and the second in \eqref{sym3}
\small
\begin{equation*}
	\mu_1\left(x,\widehat{\lda}\right)=\frac{1}{i\gamma a\left(\widehat{\lda}\right)}W\left(-x,-\widehat{\lda}\right)\widehat{\mu}_1(-x,-\widehat{\lda}),\quad \mu_2\left(x,\widehat{\lda}^*\right)=\frac{1}{i\gamma a\left(-\widehat{\lda}^*\right)}W\left(-x,-\widehat{\lda}^*\right)\widehat{\mu}_2\left(-x,-\widehat{\lda}^*\right).
\end{equation*}
\normalsize
A direct calculation shows that
\small
\begin{equation*}
	W(-x,-\widehat{\lda})=\sigma_3q_2(-x)\begin{pmatrix}
		-i\gamma & 0\\ 0&0
	\end{pmatrix}\sigma_3q_2(-x)^{-1},\, W(-x,-\widehat{\lda}^*)=\sigma_3q_2(-x)\begin{pmatrix}
		0 & 0\\ 0&-i\gamma
	\end{pmatrix}\sigma_3q_2(-x)^{-1},
\end{equation*}
\normalsize
which implies \small $\mu_1\left(x,\widehat{\lda}\right)=\frac{1}{ a\left(\widehat{\lda}\right)}\sigma_3q_2(-x)\sigma_3\begin{pmatrix}
		1\\0
	\end{pmatrix},\, \mu_2\left(x,\widehat{\lda}^*\right)=-\frac{1}{a\left(-\widehat{\lda}^*\right)}\sigma_3q_2(-x)\sigma_3\begin{pmatrix}
	0\\1
\end{pmatrix}.$ \normalsize   
Therefore, one has \small $q_1(x) =\sigma_3\left(\mu_1\left(x,\widehat{\lda}\right),\mu_2\left(x,\widehat{\lda}^*\right)\right)
	=-q_2(-x)\begin{pmatrix}
		\frac{1}{a\left(\widehat{\lda}\right)}&0 \\ 0&\frac{1}{a\left(-\widehat{\lda}^*\right)}
	\end{pmatrix}\sigma_3,$ \normalsize 
from which a direct calculation shows that 
\begin{equation}
    E_1(x)=\sigma_3E_2(-x)\sigma_3.
\end{equation}
From the direct scattering problem, we know that $\frac{\Psi_-^{(1)}\left(x,\widehat{\lda}\right)}{a\left(\widehat{\lda}\right)}=e^{-i\widehat{\lda}x}\mu_1\left(x,\widehat{\lda}\right)$.
Hence, since $a\left(\widehat{\lda}\right)\neq 0$ and $\gamma_+=\gamma_-$, from \eqref{constant_mu} (the constant $\mu_2=0$) we have
\small
\begin{align*}
	\varphi(x)  &=\sigma_3\left(\frac{\Psi_-^{(1)}\left(x,\widehat{\lda}\right)}{a\left(\widehat{\lda}\right)},\frac{N{\Psi_-^{(1)}}^*\left(x,\widehat{\lda}\right)}{a^*\left(\widehat{\lda}\right)}\right)=\sigma_3\left(e^{-i\widehat{\lda}x}\mu_1\left(x,\widehat{\lda}\right),e^{i\widehat{\lda}^*x}N\mu_1^*\left(x,\widehat{\lda}\right)\right)
    = q_1(x)
    \begin{pmatrix}
	    e^{-i\widehat{\lda}x}&0 \\ 0&e^{i\widehat{\lda}^{*}x}
    \end{pmatrix}.
\end{align*}\normalsize
It follows that,
\begin{equation}
	P(x)=\varphi(x)\left(\frac{i\gamma}{2}\right)\sigma_3\varphi(x)^{-1}\sigma_3=q_1(x)\left(\frac{i\gamma}{2}\right)\sigma_3q_1(x)^{-1}\sigma_3=E_1(x).
\end{equation}
The case $\gamma>0$ will lead to the same conclusion, one needs to use the others equations in \eqref{sym1} and \eqref{sym3} instead.
Let $\mu(x,\lda)=\1+\frac{m(x)}{\lda}+O\left(\lda^{-2}\right)$ and $a(\lda)=1+\frac{a_1}{\lda}+O\left(\lda^{-2}\right)$ be the asymptotic expansions as $\lda\to\infty$. As $\widetilde{\mu}(x,\lda)=\mu^*(-x,-\lda^*)$, it follows from (\ref{new_RHP}) that \small$-m^*(-x)=N\left(m(x)+(a_1-\rho)\sigma_3+\sigma_3(P(-x)+P(x))\right)N^{-1}$ \normalsize
Again, using the symmetries $\mu(x,\lda)=N\mu^*(x,\lda^*)N^{-1}$, $P(x)=NP^*(x)N^{-1}$ we see that $m_{12}(x)=-m_{21}^*(x)$ and $P_{12}(x)=-P_{21}^*(x)$. Then, it follows that 
\small
\begin{align*}
	u(x) =2im_{12}(x)
	& = -2i\left[N\left(m(-x)+a_1\sigma_3-\rho\sigma_3+\sigma_3P(x)+\sigma_3P(-x)\right)N^{-1}\right]_{12}^*\\
	& = 2i\left[m(-x)+P(-x)+P(x)\right]_{21}^*
	 = 2im_{21}^*(-x)+\left[P^*(x)+P^*(-x)\right]_{21}\\
	& = -2im_{12}(-x)-2i\left[P(x)+P(-x)\right]_{12}
	 =-\left(u(-x)+2i\left[P(x)+P(-x)\right]_{12}\right)\\
	& =-\left(u(-x)+2i\left[P(x)+P(-x)\right]_{12}\right)=-\widetilde{u}(-x).
\end{align*}
\normalsize
\finprf

\bibliographystyle{abbrv}
\bibliography{references}
\end{document}